\documentclass{aa}  

\usepackage{graphicx}
\usepackage{txfonts}
\begin{document}

   \title{The role of binaries in the enrichment of the early Galactic halo.}

   \subtitle{II. Carbon-Enhanced Metal-Poor Stars -- CEMP-no stars}

   \author{T.T. Hansen
          \inst{1}
          \and
          J. Andersen\inst{2,3}\and B. Nordstr{\"o}m\inst{2,3}\and
          T. C. Beers\inst{4}\and V.M. Placco\inst{4}\and J. Yoon\inst{4}\and
          L.A. Buchhave\inst{5,6}} 

   \institute{Landessternwarte, ZAH, Heidelberg University,
              K{\"o}nigstuhl 12, Heidelberg, D-69117, Germany\\
              \email{thansen@lsw.uni-heidelberg.de}
         \and
             Dark Cosmology Centre, The Niels Bohr Institute, University of Copenhagen,
             Juliane Maries Vej 30, DK-2100 Copenhagen, Denmark\\
             \email{ja@astro.ku.dk, birgitta@astro.ku.dk}
\and
    Stellar Astrophysics Centre, Department of Physics and Astronomy, Aarhus
    University, DK-8000 Aarhus C, Denmark 
\and
Department of Physics and JINA Center for the Evolution of the Elements,
University of Notre Dame, Notre Dame, IN 46556, USA\\ 
\email{tbeers@nd.edu,vplacco@nd.edujyoon4@nd.edu}            
\and
Harvard-Smithsonian Center for Astrophysics, Cambridge, MA 02138, USA
\and
Centre for Star and Planet Formation, University of Copenhagen,DK-1350 Copenhagen, Denmark \\
\email{buchhave@astro.ku.dk}
}
   \date{}

 
  \abstract
   {The detailed composition of most metal-poor halo stars has been found to be very 
     uniform. However, a fraction of 20$-$70\% (increasing with decreasing metallicity)
     exhibit dramatic enhancements in their abundances of 
     carbon -- the so-called carbon-enhanced metal-poor (CEMP) stars.
     A key question for Galactic chemical evolution
     models is whether this non-standard composition reflects that of the stellar natal
     clouds, or is due to local, post-birth mass transfer of chemically processed
     material from a binary companion; CEMP
     stars should then all be members of binary systems. }
   {Our aim is to determine the frequency and orbital parameters of binaries
     among CEMP stars with and without over-abundances of neutron-capture
     elements -- CEMP-$s$ and CEMP-no stars, respectively -- as a test of 
     this local mass-transfer scenario. This paper discusses a sample of 24 
     CEMP-no stars, while a subsequent paper will consider a similar 
     sample of CEMP-$s$ stars.} 
   {High-resolution, low-S/N spectra of the stars were obtained at roughly
     monthly intervals, over a time span of up to eight years, with the
     FIES spectrograph at the Nordic Optical Telescope. Radial velocities of 
     $\sim100$~m~s$^{-1}$ precision were determined by cross-correlation after 
     each observing night, allowing immediate, systematic follow-up of any variable object.} 
   {Most programme stars exhibit no statistically significant radial-velocity
     variation over this period and appear to be single, while four are found to 
     be binaries with orbital periods of 300$-$2,000 days and normal
     eccentricity; the binary frequency for the sample is 17$\pm$9\%. 
     The single stars mostly belong to the recently-identified ``low-C
     band'', while the binaries have higher absolute carbon abundances.} 
   {We conclude that the nucleosynthetic process responsible for the strong
     carbon excess in these ancient stars is unrelated to their binary status;
     the carbon was imprinted on their natal molecular clouds in the early Galactic ISM by an even
     earlier, external source, strongly indicating that the CEMP-no stars are likely {\it bona fide} 
     second-generation stars. We discuss potential production sites for carbon
     and its transfer across interstellar distances in the early ISM, and  
     implications for the composition of high-redshift DLA systems.}  

 
   \maketitle
%

\section{Introduction}

\begin{table*}
\caption{The sample of CEMP-no stars monitored for radial-velocity variation}
\label{tbl-1}
\centering
\begin{tabular}{lrrrrcrrrc}
\hline\hline
Stellar ID & RA (J2000) & Dec (J2000) & $V$ & $B-V$ & Ref &$\mathrm{[Fe/H]}$ &
$\mathrm{[C/Fe]}$ & $\mathrm{[Ba/Fe]}$ & Ref\\
\hline
\object{HE~0020$-$1741} & 00:22:44 & $-$17:24:28 & 12.89 & 0.94 &a&$-$4.11&$+$1.36&$<-$0.67&1 \\
\object{CS~29527$-$015}$^*$ & 00:29:11 & $-$19:10:07 & 14.26 & 0.40 &d&$-$3.55&$+$1.18&$<+$0.10&10 \\
\object{CS~22166$-$016} & 00:58:24 & $-$14:47:07 & 12.75 & 0.65 &b&$-$2.40&$+$1.02&$-$0.37&4\\
\object{HE~0219$-$1739} & 02:21:41 & $-$17:25:37 & 14.73 & 1.52 &b&$-$3.09&$+$1.90&$<-$1.39&1\\
\object{BD$+$44$^\circ$493}& 02:26:50&$+$44:57:47 &  9.22 & 0.67 &a&$-$3.83&$+$1.35&$-$0.60&2\\
\object{HE~0405$-$0526} & 04:07:47 & $-$05:18:11 & 10.72 & 0.71 &a&$-$2.18&$+$0.92&$-$0.22&1\\
\object{HE~1012$-$1540}$^*$ & 10:14:53 & $-$15:55:53 & 14.04 & 0.66 &a&$-$3.51&$+$2.22&$+$0.19&6\\
\object{HE~1133$-$0555} & 11:36:12 &$+$06:11:43 & 15.43 & 0.64 &b&$-$2.40&$+$2.20&$-$0.58&1 \\
\object{HE~1150$-$0428} & 11:53:07 & $-$04:45:03 & 15.01 & 0.76 &a&$-$3.21&$+$2.28&$-$0.44&6\\ 
\object{HE~1201$-$1512} & 12:03:37 & $-$15:29:33 & 13.79 & 0.55 &a&$-$3.92&$+$1.60&$<-$0.34&3\\
\object{HE~1300$+$0157} & 13:02:56 & $+$01:41:52 & 14.06 & 0.48 &b&$-$3.49&$+$1.31&$-$0.74&6\\
\object{BS~16929$-$005} & 13:03:30 & $+$33:51:09 & 13.61 & 0.62 &b&$-$3.27&$+$0.99&$-$0.41&3\\
\object{HE~1300$-$0641} & 13:03:34 & $-$06:57:21 & 14.80 & 0.62 &b&$-$3.14&$+$1.29&$-$0.77&8\\
\object{HE~1302$-$0954} & 13:04:58 & $-$10:10:11 & 13.96 & 0.79 &a&$-$2.25&$+$1.17&$<-$0.53&1\\
\object{CS~22877$-$001} & 13:13:55 & $-$12:11:42 & 12.16 & 0.77 &b&$-$2.71&$+$1.00&$-$0.49&5\\
\object{HE~1327$-$2326}$^*$ & 13:30:06 & $-$23:41:54 & 13.53 & 0.44 &b&$-$5.76&$+$4.26&$<+$1.46&3 \\
\object{HE~1410$+$0213} & 14:13:06 & $+$01:59:21 & 13.05 & 1.14 &b&$-$2.14&$+$1.71&$-$0.26&6\\ 
\object{HE~1506$-$0113} & 15:09:14 & $-$01:24:57 & 14.44 & 0.64 &a&$-$3.54&$+$1.65&$-$0.80&3\\
\object{CS~22878$-$027} & 16:37:36 & $+$10:22:08 & 14.41 & 0.44 &c&$-$2.52&$+$0.86&$<-$0.75&3\\
\object{CS~29498$-$043} & 21:03:52 & $-$29:42:50 & 13.63 & 1.12 &b&$-$3.75&$+$1.90&$-$0.45&7\\
\object{CS~29502$-$092} & 22:22:36 & $-$01:38:24 & 11.87 & 0.77 &b&$-$2.99&$+$0.96&$-$1.20&3\\
\object{HE~2318$-$1621} & 23:21:22 & $-$16:05:06 & 12.73 & 0.68 &a&$-$3.67&$+$1.04&$-$1.61&9\\
\object{CS~22949$-$037} & 23:26:30 & $-$02:39:58 & 14.36 & 0.79 &b&$-$3.93&$+$1.01&$-$0.77&6\\
\object{CS~22957$-$027} & 23:59:13 & $-$03:53:49 & 13.62 & 0.80 &b&$-$3.06&$+$2.13&$-$0.96&6\\
\hline
\end{tabular}
\tablebib{Photometry:(a) \citet{henden2015}; (b) \citet{beers2007}; (c)
  \citet{preston1991}; (d) \citet{norris1999}\\
Abundances:
(1) This work; (2) \citet{ito2013}; (3) \citet{yong2013}; (4)
\citet{giridhar2001}; (5) \citet{aoki2002a}; (6) \citet{cohen2013}; (7)
\citet{aoki2002c}; (8) \citet{barklem2005}; (9) \citet{placco2014a}; (10)
\citet{bonifacio2009} and M. Spite (priv. communication).}
\tablefoot{$^*$ Stars not formally qualifying as CEMP-no stars,
according to the $\mathrm{[Ba/Fe]}$ criterion.  However, see discussion of their
light-element abundance signatures in Section \ref{abundance}.}
\end{table*}

Over the past few decades, the collective effort of large-scale
spectroscopic surveys to identify and analyse very metal-poor (VMP;
$\mathrm{[Fe/H]} < -2.0$) and extremely metal-poor (EMP;
$\mathrm{[Fe/H]} < -3.0$) stars in the halo system of the Milky Way have
provided an increasingly detailed picture of the star-to-star
elemental-abundance variations that constrain the early chemical
evolution of the Galaxy \citep{cayrel2004,bonifacio2009}; for general
reviews, see \citet{beerschristlieb2005,ivezic2012,frebelnorris2015}. 

However, chemically peculiar stars exist, the abundance patterns of
which deviate markedly from those of the bulk of Population II stars.
Identifying the nature and origin of these stars should enable
identification of their progenitors, and lead to a better understanding of
the processes of nucleosynthesis and mass ejection that are responsible for 
the production of their distinctive chemical signatures. This information
will constrain the nature of the very first generations of stars, prior to
the dilution of these signatures by later mixing with the ISM of the
Galaxy, which ultimately establishes the mean abundance trends for
relatively more metal-rich stars.

A key element in this context is carbon, the first heavy element
synthesised after the Big Bang, and found with increasing frequency in
very metal-poor stars. Carbon-enhanced metal-poor (CEMP) stars were
originally identified among the VMP and EMP stars discovered in the HK
survey of Beers, Preston, \& Shectman \citep{beers1985,beers1992}, and
supplemented by a number of surveys since. The fraction of CEMP
stars rises with decreasing iron abundance, $\mathrm{[Fe/H]}$ (the
conventional metallicity indicator), which makes them particularly
important for studies of the very first stages of the formation and
evolution of the Galactic halo. 

The CEMP stars comprise a number of different sub-classes
\citep[initially defined by][]{beerschristlieb2005}. The best-populated
CEMP sub-classes are: 

{\it (i)} The CEMP-$s$ stars (indicating the presence of
$s$-process elements in addition to the carbon enhancement;
$\mathrm{[C/Fe]} > +0.7$ and $\mathrm{[Ba/Fe]} > +1.0$), the great
majority of which can be accounted for by scenarios involving the
transfer of enriched material from a binary companion that has passed
through the asymptotic giant-branch (AGB) stage of
evolution\footnote{It should be noted that, even though progress is
being made, the high fraction of CEMP-$s$ stars found in the early
Universe still poses a challenge for theoretical population synthesis
modelling \citep[e.g.,][]{izzard2009,abate2013,abate2015}.}, and 

{\it (ii)} The CEMP-no stars (indicating carbon enrichment, but no
enhancement in neutron-capture elements; $\mathrm{[C/Fe]} > +0.7$ and
$\mathrm{[Ba/Fe]} < 0.0$), the origin of which has still not been
identified with certainty. A number of lines of evidence (described in
detail below) strongly suggest association of the CEMP-no stars with the
nucleosynthesis products of the very first stars born in the Universe,
i.e., that they are bona fide second-generation stars.

\citet{aoki2002a} and \cite{ryan2005} first called attention to the apparent
contrast in the heavy-element abundance patterns among several
sub-classes of CEMP stars, and speculated about their likely origin. In
particular, \cite{ryan2005} predicted that the binary fraction of the
CEMP-no stars should be the same as for non carbon-enhanced, metal-poor
stars, if they acquired their carbon from the explosions of
previous-generation massive stars. Moreover,
\citet{carollo2012,carollo2014} found the bulk of CEMP stars in the outer halo
to be CEMP-no stars, while the CEMP-$s$ stars appear to be predominantly
associated with the inner halo -- a finding of substantial, if as yet
unclear, significance for our understanding of the formation of the
haloes of galaxies, including our own.

The present series of papers considers the origin of chemically peculiar
stars at very low metallicity, and tests whether their distinctive
abundance signatures can be accounted for by alteration of their birth
chemistry by an evolved binary companion, based on an eight-year programme 
of systematic, precise radial-velocity monitoring for larger samples of
chemically peculiar stars than have heretofore been studied in this way. 

What sets this entire project apart from earlier studies is our
emphasis on high precision; the homogeneity of the data over a time span
compatible with the periods of likely binaries in the sample; and our 
systematic follow-up of any suspected variable objects in the sample. This 
enables us to separate single stars and binaries on a rigorous, objective
star-by-star basis.

\citet{hansen2011} first showed that the enhancement of $r$-process elements 
observed in a small fraction (3-5\%) of VMP and EMP stars is {\it not
causally connected} to membership in a binary system, a conclusion that
was confirmed and further strengthened in Paper~I of this series
\citep{hansen2015b}. The present paper examines the same question for
the class of CEMP-no stars, based on similarly precise and systematic 
radial-velocity monitoring. Paper~III of this series addresses the role that
binaries play in the production of CEMP-$s$ stars, using the same
approach.  

This paper is outlined as follows. Section 2 summarises our observing strategy 
and selection of programme stars, and briefly describes the observational
techniques employed. The results are presented in Sect. 3. In Sect. 4, we
discuss the constraints imposed by these results on the origin of
CEMP-no stars, and outline the current evidence that they can be
identified with the first stars able to form in the early Galaxy and
Universe. Brief conclusions and thoughts on what can be learned from
future radial-velocity monitoring of CEMP-no stars are presented in
Sect. 5.


\section{Observing strategy, observations, and data analysis}

\subsection{Observing strategy for this project} 
\label{strategy}

Previous forays into this subject \citep[e.g.,][]{lucatello2005,starkenburg2014} 
have typically comprised a limited number of their own radial-velocity observations 
over a limited time span, sometimes combined with a literature search for
earlier data, usually small in number and necessarily heterogeneous in
origin. A Monte Carlo simulation was then used to assess the likely frequency
of typical binaries in the sample.

Our approach throughout this project is completely different, building 
on the earlier examples
of \citet{duquennoy1991}, \citet{nordstrom1997}, and \citet{carney2003}: To
obtain a sufficient number of precise radial-velocity observations in a
homogeneous system over a sufficiently long time span to detect any binaries
in the sample, even if their orbital periods might be substantially longer
than the time span of the observations (up to $\sim$2,900 days in our
case). Moreover, the observations are reduced promptly after each observing
night, in order to rapidly detect any incipient variability and schedule
follow-up observations as appropriate for each object, so the reality and
origin of the variations could eventually be assessed with confidence.  

As described in Paper~I, this strategy enabled us to follow the star
\object{HE 1523$-$0901} over several orbital revolutions and identify it
as a very low-inclination binary, despite its period of 303 days
($\sim$a year) and velocity semi-amplitude of only 0.35 km~s$^{-1}$. In
Paper~III it has also enabled us to detect several binaries with periods
of a decade or (much) more, although we 'only' followed them for
$\la$3,000 days. 

In such situations, the $\chi^2$ statistic is a much more secure
indicator of velocity variability for a given star than Monte Carlo
simulations of variations in published data, although judgement must
still be exercised in each case when reviewing the probability,
$P(\chi^2)$, that a star is truly single. For example, the low-amplitude
binary \object{HE 1523$-$0901} has $P(\chi^2) < 10^{-6}$ (Paper~I),
while the CEMP-$s$ star \object{HE 0017$-$0055}, classified as a
long-period binary in Paper~III, also exhibits an apparently similar, 
regular, short-period variation of period 385 days and amplitude 0.54
km~s$^{-1}$, followed over 8 cycles. Yet, \citet{jorissen2015}
concluded, based on other evidence, that the short-period variations 
were due to pulsations rather than binary motion (see also the discussion 
of \object{HE 1410$+$0213} below).

\subsection{Sample selection}

Our initial observing list, established in 2006, comprised a total of 23
CEMP stars with pedigrees of widely different quality, plus 17
$r$-process enhanced stars, drawn from the HK surveys of Beers, Preston,
\& Shectman \citep{beers1985,beers1992} and the Hamburg/ESO survey of
Christlieb and collaborators \citep{christlieb2008}. The initial sample of
CEMP stars also included one of the most metal-poor stars known to date,
\object{HE~1327$-$2326}, with $\mathrm{[Fe/H]} = -5.8$ \citep{aoki2006,frebel2006}. 

By 2010, the overall conclusion of the $r$-process programme had already
become clear \citep{hansen2011}, while the significance of the different
spatial distributions of the CEMP-$s$ and CEMP-no subclasses, and the
potentially different binary frequencies and origins of their carbon
excess, had assumed greater importance. At the same time, our observing
technique had been refined and amply tested as described above. From
2011, the $r$-process programme was therefore limited to sparsely
sampled long-term monitoring, while 25 likely new CEMP-no stars were
selected from the same sources and added to the regular programme. We
then also included the bright CEMP-no star BD$+$44$^\circ$493 in the
programme, the abundance pattern of which has been studied extensively
\citep{ito2013,placco2014b}. Ultimately, two of the candidate CEMP-no stars
were shown not to be sufficiently carbon enhanced to be considered CEMP
stars, and were dropped from our programme. In return, one of our candidate
CEMP-$s$ stars, to be discussed in Paper~III, turned out to be a CEMP-no
star and hence was moved to the sample discussed here.  

Our final sample of 24 CEMP-no programme stars is provided in Table
\ref{tbl-1}, which lists their $V$ magnitudes, $B-V$ colours, and
reported $\mathrm{[Fe/H]}$, $\mathrm{[C/Fe]}$, and $\mathrm{[Ba/Fe]}$
abundances (either from the literature or determined as described
below).

\subsection{Abundance information for the sample stars}
\label{abundance}

As seen from Table \ref{tbl-1}, many of our sample stars have detailed
abundance information available from high-resolution spectroscopy
obtained with 6$-$10-m telescopes (the ``First Stars'' project, e.g.,
\citealt{bonifacio2009}, is a notable exception, since that project
deliberately avoided carbon-enhanced stars). On that background, the present
project at the 2.5-m Nordic Optical Telescope (NOT) focused on establishing
the binary status of known CEMP stars. 

Only five of the stars in our sample had no Ba abundance measurement in the
published literature; one star (\object{HE~0405$-$0526}) lacked
published estimates of $\mathrm{[Fe/H]}$ and $\mathrm{[C/Fe]}$ as well.
For HE~0405$-$0526, we therefore derived estimates of $\mathrm{[Fe/H]}$
and $\mathrm{[C/Fe]}$ from a medium-resolution ($R \sim 2000$) spectrum
obtained with the SOAR Telescope (program SO2011B-002), using the n-SSPP
software pipeline described in detail by \citet{beers2014}. For this
star, as well as for the remaining stars with initially missing Ba
abundances, we have used the high-resolution spectra obtained for the
radial-velocity monitoring. These have been co-added to produce a higher
signal-to-noise ratio (SNR) spectrum, following the description in
Paper~I for co-add templates, but including only the orders containing
the Ba lines at $\lambda=4554$\,{\AA} and $\lambda=4934$\,{\AA}. 

Barium abundances or upper limits were then obtained from 
spectral synthesis of these spectra, using the 2014 version of MOOG and
the line list retrieved from the VALD database \citep{kupka1999},
including hyperfine splitting and isotopic shifts. The
\citet{asplund2009} solar abundances have been assumed. The co-added spectra may
have slightly broadened lines, due to the initial correlation when creating
this spectrum, as described in Paper~I, which could influence the abundances
derived from these and result in a higher abundance estimate. We have
taken this into account when estimating the error on the derived abundances.

For three of the stars we considered, only an upper limit on the Ba
abundance could be derived; however, all of these firmly classify the stars as
CEMP-no stars. For HE~0405$-$0526 and HE~1133$-$0555, we derive Ba abundances
of $\mathrm{[Ba/Fe]}=-0.22$ and $\mathrm{[Ba/Fe]}=-0.58$, respectively, with an
estimated error of 0.3~dex, also classifying them as CEMP-no stars. 

For the three stars CS~29527$-$015, HE~1012$-$1540, and HE~1327$-$2326, the 
derived Ba abundances or upper limits ($\mathrm{[Ba/Fe]} < +0.10$,
$\mathrm{[Ba/Fe]} = +0.19$, and $\mathrm{[Ba/Fe]} < +1.46$, respectively) do
not qualify them as CEMP-no stars, based on the formal definition
($\mathrm{[Ba/Fe]} < 0$). 
However, enhancement of the light elements has also been found to be 
associated with the CEMP-no sub-class; see the recent discussion by
\citet{norris2013b}. For HE~1012$-$1540, \citet{cohen2013} found
$\mathrm{[N/Fe]} = +1.25$, $\mathrm{[O/Fe]} = +2.14$, $\mathrm{[Na/Fe]} =
+1.02$, and $\mathrm{[Mg/Fe]} = +1.38$, while \citet{frebel2008} derived $\mathrm{[N/Fe]}
= +4.53$, $\mathrm{[O/Fe]} = +3.68$, $\mathrm{[Na/Fe]}= +2.17$ and
$\mathrm{[Mg/Fe]} = +1.67$ for HE~1327$-$2326. \citet{frebel2008} also
obtained a relatively high Sr abundance for this star, $\mathrm{[Sr/Fe]} =
+0.98$. An absolute carbon abundance of $A$(C) $= \log\epsilon\,(C) = 6.06$
was reported for CS~29527$-$015 by \citet{bonifacio2009},
also pointing toward a CEMP-no classification for this star, but further
observations are needed to confirm this.

\subsection{Radial-velocity observations and data reduction}

The observations, data reduction, and analysis procedures were those described 
in Paper~I of this series, to which we refer the interested reader for details; 
only a brief summary is given here. Weather allowing, our programme stars
were observed at roughly monthly intervals with the FIES spectrograph at 
the 2.5-m NOT on La Palma, Spain. The spectra cover a 
wavelength range of 3640\,{\AA} to 7360\,{\AA} at a resolving power of 
$R \sim46,000$, and have an average SNR of 10. For obvious reasons, the stars 
added to the programme in 2010 were observed for shorter spans and, coupled 
with adverse spring weather conditions later, were observed less completely 
than stars from the initial sample. 

Reductions and multi-order cross-correlation with a template spectrum
were performed with software developed by L. Buchhave. The template
spectrum used for a given star was either: The spectrum with maximum SNR
(``strongest''); a co-added spectrum constructed from all the best
spectra for the star (``Co-add''); a synthetic spectrum consisting of
delta functions (``Delta''); or, finally, co-added spectra of the bright
CEMP-no stars BD$+$44$^\circ$493 or HE~0405$-$0526. The template
used for each star is identified in Table \ref{tbl-2}.

Our error definitions and error analysis are also described in Paper~I.
Note especially that the standard deviations given in Tables \ref{tbl-2}
and \ref{tbl-3} are the standard deviation of the radial velocity
observations for each star, not standard errors of the mean. 

The individual standard stars observed on this programme, along with their 
derived mean heliocentric radial velocities (RV) and standard deviations
($\sigma$), are listed in Table~2 of Paper~I in this series; typical 
standard deviations are 40~m~s$^{-1}$ over a time span of $\sim$2,800 
days. The average difference of our mean velocities from the standard 
values is 73~m~s$^{-1}$ with a standard deviation of 69~m~s$^{-1}$, 
demonstrating that our results are not limited by the stability of the 
FIES spectrograph.

\section{Results}
\begin{table*}
\caption{Star name; number of NOT observations; template used; mean
  heliocentric radial velocity and standard deviation; time span; variability criterion $P(\chi^2)$; and
  binary status.}  
\label{tbl-2}
\centering
\begin{tabular}{lrlrrrlc}
\hline\hline
Stellar ID & Nobs & Template & RV mean & $\sigma$ & $\Delta$T & $P(\chi^2)$ &Binary \\
           &      &          &(km~s$^{-1}$)&(km~s$^{-1}$)& (Days)&  &  \\
\hline
\object{HE~0020$-$1741}   &  9& Co-add            &  $+$9.039 & 0.212 & 1066&0.122&No\\
\object{CS~29527$-$015}   &  6& BD$+$44$^\circ$493 & $+$47.122 & 0.425 & 1064&0.992&No\\
\object{CS~22166$-$016}   &  8& Co-add            &$-$210.504 & 0.803 & 1034&0.255&No\\
\object{HE~0219$-$1739}   & 15& Co-add            &$+$106.689 & 5.090 & 2207&0.000&Yes\\
\object{BD$+$44$^\circ$493}& 18& Co-add            &$-$150.084 & 0.051 & 1298&0.927$^*$&No\\
\object{HE~0405$-$0526}   & 13& Co-add            &$+$165.657 & 0.039 &  904&0.993&No\\
\object{HE~1012$-$1540}   &  8& HE~0405$-$0526    &$+$226.052 & 0.207 &  802&0.988&No\\
\object{HE~1133$-$0055}   &  9& HE~0405$-$0526    &$+$270.632 & 0.336 & 2217&0.707&No\\
\object{HE~1150$-$0428}   & 13& Strongest         & $+$48.042 & 8.350 & 2220&0.000&Yes\\
\object{HE~1201$-$1512}   &  5& Delta             &$+$239.450 & 1.854 &  42 &0.582&No\\
\object{HE~1300$+$0157}   &5 & BD$+$44$^\circ$493  & $+$74.494 & 0.583 &  412&0.122$^*$&No\\
\object{BS~16929$-$005}   &7 & Delta              & $-$50.627 & 0.474 &  885&0.912$^*$&No \\
\object{HE~1300$-$0641}   &  2& HE~0405$-$0526    & $+$68.846 & 0.114 &  386&0.827&No\\  
\object{HE~1302$-$0954}   &  3& BD$+$44$^\circ$493 & $+$32.538 & 0.039 &  386&0.968&No\\
\object{CS~22877$-$001}   &15& Co-add             &$+$166.297 & 0.111 & 2923&0.737$^*$&No\\
\object{HE~1327$-$2326}   &  9& Delta             & $+$64.343 & 1.170 & 2577&0.545&No\\
\object{HE~1410$+$0213}   & 23& Co-add            & $+$81.140 & 0.180 & 3006&0.196&No\\
\object{HE~1506$-$0113}   & 10& Delta             & $-$81.467 & 2.772 &  493&0.000$^*$&Yes \\
\object{CS~22878$-$027}   & 7& Delta              & $-$90.870 & 0.768 & 1034&0.775$^*$&No\\ 
\object{CS~29498$-$043}   & 15& Delta             & $-$32.488 & 0.701 & 2603&0.967&No\\
\object{CS~29502$-$092}   & 20& Co-add            & $-$67.160 & 0.173 & 2603&0.616$^*$&No\\
\object{HE~2318$-$1621}   &  7& BD$+$44$^\circ$493 & $-$41.698 & 0.279 & 1034&0.172&No\\
\object{CS~22949$-$037}   & 7& BD$+$44$^\circ$493  &$-$125.560 & 0.269 &  765&0.894$^*$&No\\
\object{CS~22957$-$027}   & 18& Co-add            & $-$67.305 & 5.736 & 1568&0.000&Yes\\ 
\hline
\end{tabular}
\tablefoot{$^*$ Additional literature data included in $P(\chi^2)$, see
  section \ref{statistics} and Table \ref{tbl-3}.}
\end{table*}

The results of our radial-velocity observations for our CEMP-no stars
are summarised in Table \ref{tbl-2}, which lists the number of
observations (Nobs), the cross-correlation templates employed, the mean
heliocentric radial velocity (RV mean) and standard deviation ($\sigma$)
over the time span covered by our own observations ($\Delta$T), and the binary
status of each programme star. The individual observed heliocentric radial
velocities are listed in Appendix A, together with the Julian dates of the
observations and the corresponding internal errors.  

As for the $r$-process-element enhanced stars discussed in Paper~I, the
external standard deviations of the observed velocities vary from $<~100$
m~s$^{-1}$ for the bright targets, dominated by centering and guiding errors,
to $\sim1$ km~s$^{-1}$ for the fainter and/or metal-poor targets with low-SNR
spectra.

\begin{table*}
\caption{Additional data for the $P(\chi^2)$ calculations; star name, number
  of additional observations, total time span covered for the star (including NOT
  observations), and reference.}
\label{tbl-3}
\centering
\begin{tabular}{lrrl}
\hline\hline
Stellar ID & Nobs & $\Delta$T$_{tot}$ & ref\\
\hline
\object{BD$+$44$^\circ$493}&4&2401&\citet{starkenburg2014}\\
\object{HE~1300$+$0157}&6&838&\citet{starkenburg2014}\\
\object{BS~16929$-$005}&5&885&\citet{starkenburg2014}\\
\object{CS~22877$-$001}&3&4488&\citet{roederer2014}\\
\object{CS~22878$-$027}&6&1034&\citet{starkenburg2014}\\
\object{CS~29502$-$092}&3&2603&\citet{starkenburg2014}\\
\object{CS~22949$-$037}&3&3003&\citet{starkenburg2014}, \citet{roederer2014}\\
\hline
\end{tabular}
\end{table*}

\subsection{Comparison with published radial velocities for the constant stars}

\begin{figure}
\resizebox{\hsize}{!}{\includegraphics{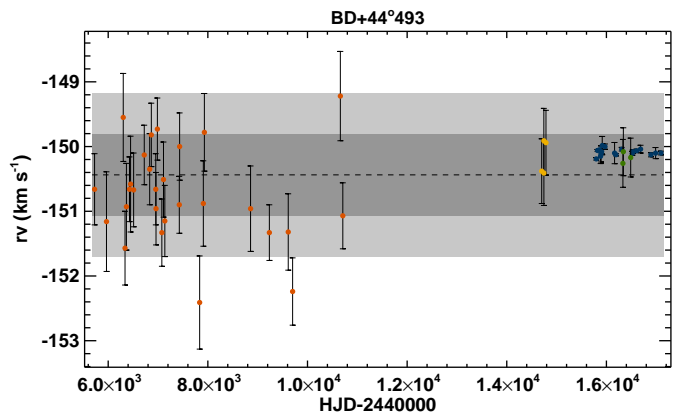}}
\caption{Radial-velocity history of BD$+$44$^{\circ}$493 vs. time.
$Blue:$ This work, $green:$ \citet{carney2003}, $yellow:$
\citet{ito2013}, and $red:$ \citet{starkenburg2014}. Dashed line: Global
mean radial velocity; grey shaded areas: 1$\sigma$ and 2$\sigma$ regions around the
global mean.} 
\label{BD44_493}
\end{figure}

Radial-velocity observations for 16 of the programme stars without obvious
velocity variations have been published previously, and may help to extend the
time span of the data and strengthen our conclusion on their (non-) binary
nature. We have searched the literature for velocities of the non-binary stars
in our sample, based on moderate- to high-resolution spectra by authors who
claim estimated velocity errors of $\sim$1~km~s$^{-1}$ or better. Table
\ref{tbl-B1} in Appendix B summarises the results of this exercise. Note
that these data are typically based on a single observation (or the mean
of several closely spaced observations). In each case, the total time
span given includes the oldest measurement that we consider reliable. 

It is remarkable that {\it no} velocity in this comparison sample
deviates from our own mean radial velocity by more than can be accounted
for by the published errors, especially since the data was obtained with
different spectrographs and were reduced and measured independently. As
seen from the table, the observed time span for many of our programme
stars now cover on the order of a decade or more. Clearly, if any of
the stars we consider to be single are in fact members of a binary system,
they are likely to have extremely long periods (but see discussion
below).

An illustrative case is the bright star BD$+$44$^\circ$493, which was
also observed by \citet{starkenburg2014}. As seen from Table~\ref{tbl-2}
and \ref{tbl-B1}, the radial velocities derived by
\citet{starkenburg2014} are in excellent agreement with our derived
velocities for both this and other constant stars (all except
CS~29502-092 are within the stated 1-$\sigma$ error bars; and
CS~29502-092 agrees within 2$\sigma$). Radial velocities for
BD$+$44$^\circ$493 were also reported by \citet{carney2003} and
\citet{ito2013}, whose measured mean velocities of $-150.64$~km~s$^{-1}$
and $-150.15$~km~s$^{-1}$, respectively, are both in excellent agreement
with our result. The Carney et al. data are particularly valuable, since
they enlarge the three-year time span of our own data by 13 years (1984
to 1997). Together, these four data sets contain 53 measurements
spanning a total of 11368 days ($\sim 31$ years) with a standard
deviation of 0.63 km~s$^{-1}$. Figure~\ref{BD44_493} shows all of the radial
velocity data for BD$+$44$^{\circ}$493 as a function of time. Clearly, this
star is single. 

Among the sources of literature data in Table \ref{tbl-B1} (Appendix B), 
the only modern series of precise velocities with significant overlap with
our programme are those by \citet{starkenburg2014} and \citet {roederer2014}, 
whose velocity zero-points, moreover, agree with ours to within $\sim$100
m~s$^{-1}$. We have therefore included them in our $\chi^2$ calculations 
as detailed in Table \ref{tbl-3}, whenever they could make a significant
contribution to our results for the single stars.

\subsection{Assessment of velocity variability} 
\label{statistics}

As discussed above, the data included here were obtained in a systematic
manner and are of homogeneous quality, including the selected sources
listed in Table \ref{tbl-3}. In this situation, the $\chi^2$ statistic is
the most objective and sensitive indicator of velocity variability as
noted, e.g., by \citet{carney2003}. 

In order to assess the binary status of our sample stars, we use the 
standard $\chi^2$ criterion for velocity variability and compute the 
probability, $P(\chi^2)$, that the radial velocity is constant, based on the 
dispersion of the observed velocities vs. their associated internal errors
(see Paper~I for definitions). A ``floor error'' term of $\sim$100~m~s$^{-1}$ 
is added in quadrature to the internal error in order to account for external 
error sources such as variations in temperature or air pressure, atmospheric 
dispersion at high airmass, slit positioning, or imperfect guiding.  

Observations correlated with the ``Delta'' template presented special
difficulties, due to the small number of useful lines in each of few
usable orders in our low-signal spectra. This situation may lead to
large stochastic fluctuations in the internal mean error of individual
velocities, so that velocities with normal uncertainty, but unusually
small or large internal error estimates, completely dominate the overall
value of $\chi^2$ or, conversely, are effectively ignored. This is
exacerbated in the limiting case of faint and/or extremely metal-poor
stars at the limit of what is possible with a 2.5-m telescope, when
observations obtained under poor conditions cause the correlation with a
stellar spectrum to fail altogether. 

Alternative attempts to estimate an average overall internal error were found
to give results that were effectively biased towards either too small or too
large errors. However, an average of the (large) mean of all internal errors
and the (smaller) error of a mean observation of average weight was found to
work reliably, and was adopted in such cases. 

Adding the \citet{starkenburg2014} and \citet{roederer2014} data listed in 
Table \ref{tbl-3}, we then proceeded to calculate $P(\chi^2)$ for potentially 
single stars. The results, listed in Table \ref{tbl-2},
demonstrate that the great majority, 20 of our 24 programme stars,
have constant radial velocities over the eight-year period of monitoring
(i.e., they appear to be single), while four exhibit obvious
radial-velocity variations that are clearly due to orbital motion.  

However, limiting cases exist in which low-level variability may be
interpreted as due to either motion in a nearly face-on binary orbit as
in \object{HE~1523$-$0901}, or to pulsations like those in
\object{HE~0017$+$0055}, as discussed in Sect. \ref{strategy}. Since so
very nearly face-on binary orbits ($i\la 1 \degr$) must be exceedingly 
rare, finding even one in such a modest sample raises immediate suspicions
about its plausibility, such as discussed for the case of
\object{HE~1410$+$0213} in Sect. \ref{binaries} below. 

This finding raises another cautionary note concerning the use of Monte
Carlo simulations to assess binary frequencies from samples of even
accurate, but scattered radial-velocity observations: The limit to the
estimated frequency of long-period binaries in the sample may be set by
the presence of low-amplitude pulsations rather than merely by 
observational errors, if orbital motion is considered to be the only 
possible cause of small velocity variations in the error budget.

\subsection{Binaries in our sample of CEMP-no stars}
\label{binaries}

\begin{figure}
\resizebox{\hsize}{!}{\includegraphics{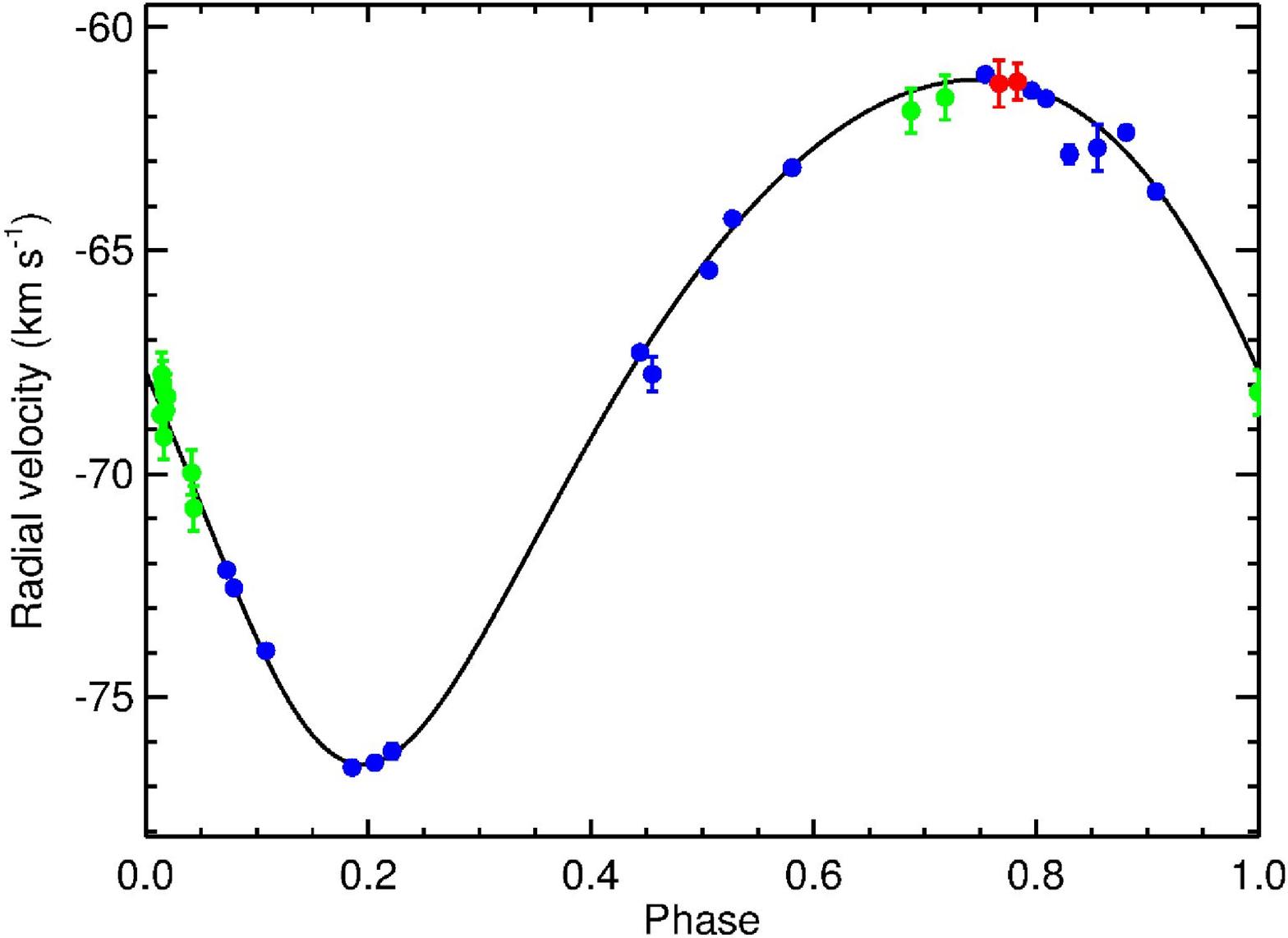}}
\resizebox{\hsize}{!}{\includegraphics{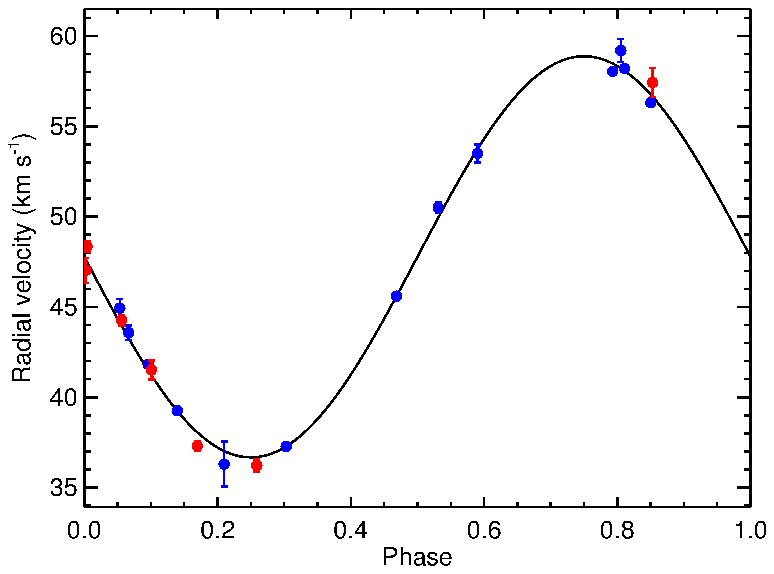}}
\resizebox{\hsize}{!}{\includegraphics{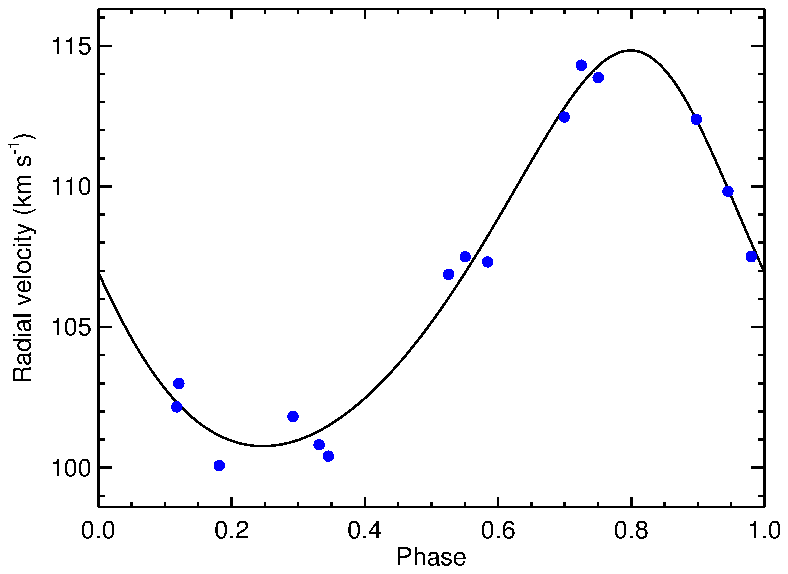}}
\caption{Orbit solutions for three of the four binaries found among our programme stars. 
Top: CS22057$-$027, middle: HE~1150$-$0428, and bottom: HE~0219$-$1739. Blue:
This work; red:\citet{starkenburg2014}; green: \citet{prestonsneden2001}} 
\label{fig:orbit1}
\end{figure}

\begin{table*}
\caption{Orbital parameters for three of the binary systems identified in the sample}
\label{tbl-4}
\centering
\resizebox{\textwidth}{!}{
\begin{tabular}{lcccc}
\hline\hline
Parameter & \object{HE~0219$-$1739} & \object{HE~1150$-$0428} 
& \object{CS~22957$-$027}\\
\hline
Period (days)        & 1802.5$\pm$5.0 &289.7$\pm$0.1&1080.0$\pm$0.8 \\   
$T_0$ (HJD)          & 2455981.4$\pm$1.7 &2455759.3$\pm$0.3 &2455660.0$\pm$1.6 \\   
$K$ (km~s$^{-1}$)     &7.032$\pm$0.022&11.102$\pm$0.043&7.661$\pm$0.039 \\ 
$\gamma$ (km~s$^{-1}$)&$+$106.809$\pm$0.016&$+$47.779$\pm$0.035 &$-$67.500$\pm$0.026\\  
$e$                  & 0.162$\pm$0.004 &0.000$\pm$0.000 &0.193$\pm$0.007 \\   
$\omega$~$\deg$      & 29.8$\pm$1.0 & 0.0$\pm$0.0 & 155.2$\pm$1.6 \\   
$a$sin $i$ (R$_{\sun}$)& 247.2$\pm$1.0 & 63.56$\pm$0.09 & 160.5$\pm$0.4 \\ 
f(m) (M$_{\sun}$) & 0.062$\pm$0.004 & 0.041$\pm$0.002 & 0.048$\pm$0.003  \\  
$\sigma$ (km~s$^{-1}$)& 0.867  & 0.670     &  0.454     \\  
$\Delta$T/P (total)& 1.2 & 7.7 & 4.7  \\ 
$R_{Roche}$ (R$_{\sun}$, M$_1$ = 0.8 M$_{\sun}$, M$_2$ = 0.4 M$_{\sun}$)& 121 & 29  & 73      \\   
$R_{Roche}$ (R$_{\sun}$, M$_1$ = 0.8 M$_{\sun}$, M$_2$ = 1.4 M$_{\sun}$) & 442 & 131 & 310 \\ 
\hline
\end{tabular}
}
\end{table*}

Four stars in our sample exhibited clear orbital motion during the monitoring period:
\object{HE~0219$-$1739}, \object{HE~1150$-$0428}, \object{HE~1506$-$0113}, and
\object{CS~22957$-$027}. We have determined orbits for three of these systems,
combining our radial velocities with earlier, published data. The final
orbital parameters for these systems are listed in Table~\ref{tbl-3}, and the
radial-velocity curves are shown in Fig.~\ref{fig:orbit1}, including the
literature data and total time spans as noted below. For the
\citet{starkenburg2014} data (red points in Fig.~\ref{fig:orbit1}), 
we find an offset of only
$\sim18$ m~s$^{-1}$ between our data and theirs, based on six  
constant stars in common. For the data from \citet{prestonsneden2001}
(green points in Fig.~\ref{fig:orbit1}), we have applied a correction of
$\sim1.84$ km~s$^{-1}$ to achieve consistency with our velocities. 

For \object{HE~1506$-$0113}, we could combine our ten observations with 
the seven measurements by \citet{starkenburg2014}, which range from $\sim -$80 
km~s$^{-1}$ to $-$90 km~s$^{-1}$, in good agreement with our own results for
this star (see Appendix). Combining them led to a plausible low-amplitude
orbit with $P~\sim$840 d and modest eccentricity, although the data were
insufficient to determine the orbital parameters with confidence. However, the
four constant velocities of $\sim -137$ km~s$^{-1}$ from high-resolution VLT/UVES spectra 
reported by \citet{norris2013a}, spanning $\sim$1 month in 2008, did not at all 
fit anything like this orbit, but could be reconciled with a plausible highly-eccentric
orbit of $P~\sim$950 d. Nonetheless, the resulting mass function of
$f(m)~\sim~0.5~M_{\sun}$ implied an impossibly high mass for the unseen
companion, so this orbit was also rejected. 
 
\begin{figure}
\resizebox{\hsize}{!}{\includegraphics{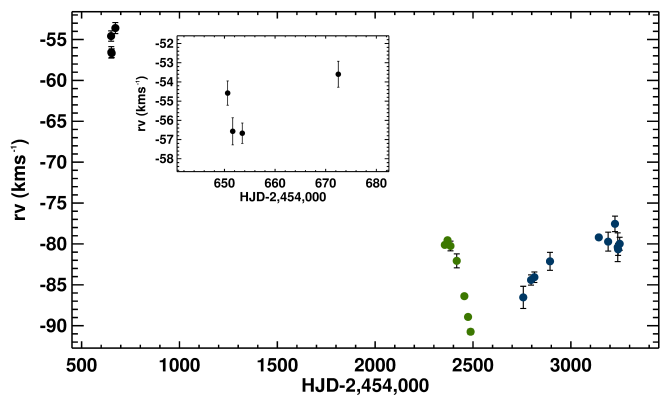}}
\caption{Radial-velocity data for HE~1506$-$0113. Blue: this work; red:
  \citet{starkenburg2014}; and black: re-analysed UVES spectra from
  \citet{norris2013a}. The inserted box is an enlargement of the RVs obtained from the UVES spectra.}
\label{fig:HE1506}
\end{figure}

The large span of the observed velocities led us to re-examine the four 
archival UVES spectra more closely, and determine fresh radial velocities from 
them. The -- very different -- results are listed in Table \ref{tbl-5} and
plotted as a function of time in Figure \ref{fig:HE1506}. These new velocities
are much {\it higher} than those reported earlier; they clearly vary over the
short observing period ($P(\chi^2)$ = 0.001), but delineate a shallow  velocity
{\it minimum} instead of the expected maximum, as is shown in the small inserted
box in Figure \ref{fig:HE1506}.  

Alternative explanations for this recalcitrant behaviour include a triple system,
or a possible misidentification with a star of similar brightness 15$\arcsec$ 
SE of \object{HE~1506$-$0113} itself, but solving this puzzle will require
several more years of continued monitoring. In conclusion, and despite
persistent effort, we have not been able to compute an orbit for this star,
although it is clearly at least a binary.

\begin{table}
\caption{Heliocentric radial velocities for HE~1506$-$0113 from UVES spectra.}
\label{tbl-5}
\centering
\begin{tabular}{lrr}
\hline\hline
HJD & RV & RV$_{err}$\\
\hline
2454650.630731 & $-$54.58 & 0.631\\
2454651.628669 & $-$56.57 & 0.705\\
2454653.519178 & $-$56.67 & 0.530\\
2454672.486795 & $-$53.60 & 0.674\\
\hline
\end{tabular}
\end{table}

Table \ref{tbl-4} also lists the Roche-lobe radii for the
present secondaries in the three binary systems for which we have computed
orbits, calculated following the procedure described in Paper~I and assuming
present primary masses of 0.8~M$_{\odot}$ and secondary masses of 0.4 and
1.4~M$_{\odot}$, respectively. These are  the lower and upper limits to the mass of a white
dwarf (WD), which would be the likely remnant if the CEMP-no star were
polluted by an initially more massive AGB star. As can be seen from Table \ref
{tbl-4}, for the maximum WD mass (1.4~M$_\odot$), all of the systems could have
accommodated an AGB star (R $\sim$ 200~R$_\odot$), while for the lowest
WD masses, the majority of the Roche-lobes would be too small.\\

\noindent {\it Notes on individual stars:} \\

Both \citet{prestonsneden2001} and \citet{starkenburg2014} found
significant variations in the velocity of \object{CS~22957$-$027}, but
were unable to determine the orbital period securely. Our own data did
so, as kindly confirmed by Dr. G.W. Preston (priv. comm.), with the
earlier data improving the period determination. \citet{starkenburg2014}
also reported large velocity variations in \object{HE~1150$-$0428} and
\object{HE~1506$-$0113}, but the data were too sparse to determine an
orbital period. Our own, more complete data determine the orbit for
\object{HE~1150$-$0428} securely (see Fig.~\ref{fig:orbit1}),  
while we have not been able to derive an acceptable orbital solution 
for HE~1506$-$0113, as discussed above.
 
\object{HE~1410$+$0213} presented special difficulties, despite the large
number of good observations (Table~\ref{tbl-2}). With $P(\chi^2)~< 10^{-4}$, its
velocity is certainly variable, and some apparently systematic trends during
individual years led us to suspect that it was a(nother) nearly face-on
spectroscopic binary. Eventually, we were able to derive a plausible circular
orbit with a period of $\sim$330 days and a velocity amplitude of $\sim$0.28
km~s$^{-1}$. This, however, required an inclination of less than $1\degr$,
again highly unlikely in such a small sample of stars. Moreover, a later
observation did not fit the putative orbit. We were thus faced with a similar
problem as for the CEMP-$s$ star \object{HE~0017$+$0055}, discussed in
Sect.~\ref{strategy} and in Paper~III.

After much analysis, we conclude that \object{HE~1410$+$0213} is
most probably single, but exhibits low-amplitude pulsations of the kind
described by \citet{riebel2010} for C-rich variables in the LMC. Adding a
velocity 'jitter' of 150~m~s$^{-1}$ then leads to the value of $P(\chi^2) =
0.196$ given in Table~\ref{tbl-2}.

\citet{mattsson2015} describes how a combination of a 
strong stellar wind with such pulsations might enhance the mass loss
from such an intrinsically bright EMP giant or AGB star. This could
explain why the absolute carbon abundance of \object{HE~1410$+$0213},
places it in the ``high-C band'' of \citet{spite2013}, discussed further
below.

Finally, \citet{bonifacio2009} suggested that \object{CS~29527$-$015} is
a double-lined spectroscopic binary, due to asymmetries in the
absorption lines detected in their high-resolution, high(er)-SNR UVES
spectra. We find no evidence for this in our observations of the
star, and with $P(\chi^2) = 0.992$, there is also no evidence for any
binary motion, but additional high-resolution spectra may be required to
settle the issue.

\section{Discussion}

Our sample of CEMP-no stars comprises 4 binaries and 20 single stars, 
We thus derive a binary frequency of 17$\pm$9\%, identical to that found for
the $r$-process-enhanced stars in Paper~I of this series
(18\%$\pm$11\%)\footnote{Note that we have corrected the error on this
  frequency reported in Paper~I, which was incorrectly stated to be
  $\pm$6\%.}, and to the 16$\pm$4\% binary frequency found by
\citet{carney2003} in their study of 91 metal-poor ($\mathrm{[Fe/H]\leq
  -1.4}$) field red giants. 

\subsection{Binary parameters and their implications}

\begin{figure}
\resizebox{\hsize}{!}{\includegraphics{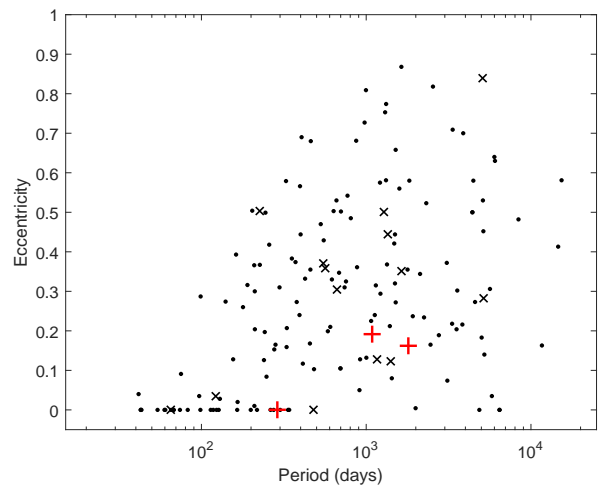}}
\caption{Period - eccentricity distribution for the binary systems in our
sample (red plus signs), compared to literature data, represented by black dots: Population 
I cluster giants (\citealt{mermio2007,mathieu1990}); and black crosses: Population I 
giants \citep{carney2003}. }
\label{fig:PE}
\end{figure}

The four binary systems in our sample exhibit a similar combination of
periods and eccentricities to those found by \citet{carney2003} for
metal-poor field red giants and dwarfs (see their Fig. 5), for
Population I cluster giants by \citet{mermio2007}, and by us for the
$r$-process-enhanced VMP and EMP stars discussed in Paper~I
\citep{hansen2015b}. Figure~\ref{fig:PE} shows the period -- 
eccentricity distribution for our binary stars (red plus signs),
compared to the literature data from \citet{mermio2007},
\citet{mathieu1990} (black dots), and \citet{carney2003} (black crosses). 
The properties of our binary stars are thus completely indistinguishable from 
those of chemically normal population I and II giants, with
\object{HE~1150$-$0428} defining a cutoff period for circular orbits of old giant 
stars of $\ga$ 300 days, presumably due to the long time available for tidal circularisation. 

This simple observational finding indicates that membership of a binary system 
has played no causal role in generating the dramatic carbon excess of the CEMP-no 
stars and rules out the local mass-transfer scenario for its origin. Instead,
the excess carbon must have been produced elsewhere and implanted across
interstellar distances into the natal cloud of the star we observe today -- a
process that is not included in standard models of the early chemical
evolution of galaxy haloes. 

\subsection{Carbon bands}

\begin{figure}
\resizebox{\hsize}{!}{\includegraphics{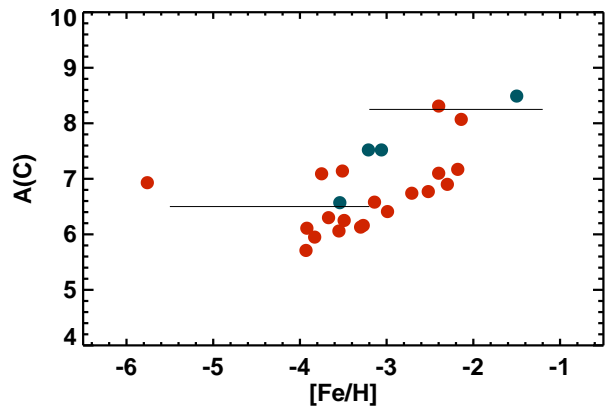}}
\caption{Absolute carbon abundances, $A(\rm C)=\log\epsilon\,(C)$, 
of our programme stars, as a function of metallicity, [Fe/H]. Blue
dots represent recognised binaries in our sample; red dots represent
non-binaries. The high-C and low-C bands of
\citet{spite2013} are indicated by the horizontal lines.} 
\label{Cband}
\end{figure}

\citet{spite2013} first suggested the existence of two bands in the absolute
carbon abundances of CEMP stars -- a high-C band at $A(\rm C) \sim
8.25$ and a low-C band at $A(\rm C) \sim 6.5$. Stars with carbon
abundances on the lower band are more metal poor
($\mathrm{[Fe/H]}\lesssim -3.0$)  and mostly
of the CEMP-no variety, while those on the higher band are relatively more
metal rich ($\mathrm{[Fe/H]}\gtrsim -3.0$), and primarily CEMP-$s$ stars.
They suggested that stars on the high-C band are members of binary
systems that have been polluted by a companion AGB star, i.e., their
carbon enhancement is {\it extrinsic}, whereas the stars on the low-C
band were born from an ISM that had already been polluted by a
(presumably) high-mass progenitor, i.e., the carbon enhancement is thus
{\it intrinsic} to the star. 

The existence of these two bands for CEMP stars has been confirmed with
larger samples by \citet{bonifacio2015} and
\citet{hansen2015a}. Both of these samples find CEMP-no stars with 
carbon abundances placing them on the high-C band, but neither of these
papers had sufficient information available to address the binary nature of their
stars, and thereby more fully test the hypothesis for the astrophysical
origin of the bands. Figure \ref{Cband} shows the absolute carbon
abundances of the stars in our sample as a function of metalicity; blue
dots represent the detected binary stars and red dots represent the
single stars. 

As can be seen from inspection of Figure \ref{Cband}, three of our CEMP-no
stars (HE~0219$-$1739, HE~1133$-$0555, and HE~1410$+$0213) have carbon
abundances corresponding to the high-C band. HE~0219$-$1739 is a binary
with ample space to accommodate a companion at the AGB stage in its
Roche lobe for the assumed minimum and maximum masses (0.4~M$_{\odot}$
and 1.4~M$_{\odot}$) of the putative white dwarf remnant (Table
\ref{tbl-3}). HE~1410$+$0213 and HE~1133$-$0555 are both most likely
single stars, but HE~1410$+$0213 is pulsating (see discussion above), which
could enhance any mass loss by a strong stellar wind.  

Two of the more metal-poor binary stars have carbon abundances that lie
in the transition area between the two bands, around $\mathrm{[Fe/H]}
\sim -3.0$, where both CEMP-no and CEMP-$s$ stars can be found. These two
stars are HE~1150$-$0428 and CS~22957$-$027 (see Fig. \ref{fig:orbit1}),
the former having a circular orbit with a period of $P$ = 290 days, and the
latter having a slightly eccentric orbit with a period of $P$ = 1080 days. Only
one of the stars with an absolute carbon abundance clearly on the low
band is found to be in a binary system, HE~1506$-$0113. This star also
exhibits high Na and Mg abundance ratios ($\mathrm{[Na/Fe]} = +1.65$ and
$\mathrm{[Mg/Fe]} = +0.89$; \citealt{yong2013}), signatures that are also
found for many other CEMP-no stars \citep{norris2013b,hansen2015a}. It is
worth noting that HE~1150$-$0428 also exhibits a high Na abundance
$\mathrm{[Na/Fe]} = +1.31$, but normal Mg abundance $\mathrm{[Mg/Fe]} = +0.36$
\citep{cohen2013}. For CS~22957$-$027, \citet{cohen2013} report a high
Na abundance ratio ($\mathrm{[Na/Fe]} = +0.80$), but a low Mg ratio
($\mathrm{[Mg/Fe]} = +0.11$), while \citet{aoki2002c} obtain a higher Mg
ratio ($\mathrm{[Mg/Fe]} = +0.69$).

It is clearly highly desirable to obtain further information on the
possible binary nature of the other six CEMP-no stars with carbon
abundances on the high-C band, found in previous work
\citep{bonifacio2015,hansen2015a}, and expand the sample of such stars.
Should the majority turn out to be members of binary systems (rather
unlikely in view of our present results), and in particular if there are
signs that mass transfer has occurred, this would lend support to the
existence of AGB stars that produce little if any $s$-process elements. 
A new nucleosynthesis process might then need to be invoked, or our 
understanding of the operation of the $s$-process at low metallicity 
must be improved. If these stars are also found to be single, another, 
distant production site must be invoked.

\subsection{Evidence that CEMP-no stars are indeed second-generation
stars}

Our primary result from this paper confirms (with much-improved
statistics) that mass transfer from a binary companion is not a
necessary condition to account for the distinctive elemental-abundance
patterns of CEMP-no stars. This is already strong evidence that they may
indeed be bona fide second-generation stars, formed from an ISM polluted
by a previous (possibly first-generation) population of stars. However,
this is not the only evidence for this association. Here we briefly
summarise the present observational constraints on this hypothesis. 

\begin{itemize}

\item{{\bf The increased frequency of CEMP stars at low metallicity}:  It has been 
recognised for over a decade that the relative numbers of CEMP stars
(compared to C-normal stars) increases dramatically as $\mathrm{[Fe/H]}$
declines from $\mathrm{[Fe/H]} = -2.0$ to the most Fe-poor star known
(SMSS~J0313-6708, with $\mathrm{[Fe/H]} \lesssim -7.8$;
\citealt{keller2014,bessell2015}). Recent large samples of VMP and EMP
stars from SDSS/SEGUE, e.g., \citet{lee2013}, have reinforced this
result based on the thousands of CEMP stars found in this survey. It has
also been shown that, at the lowest metalicities, the CEMP-no stars are
the dominant sub-class of CEMP stars \citep{aoki2007,norris2013b}.
Indeed, when limited to the sample of recognised or likely CEMP-no
stars, the derived frequencies increase by 5-10\% relative to CEMP stars
when considered as a single class \citep{placco2014a}. It has also
been argued by \citet{bromm2003} and \citet{frebel2007a} that high C (and
O) abundances facilitate low-mass star formation in the early Universe,
resulting in the birth of long-lived C-enhanced stars.} \\

\item{{\bf The dominance of CEMP-no stars among C-enhanced stars at the lowest metallicity}: 
\citet{aoki2007} found a clear difference in the metallicity
distributions of CEMP-no and CEMP-$s$ stars -- the CEMP-no stars in
their sample primarily occupy the lowest metallicity range (and
are the only C-enhanced stars found below $\mathrm{[Fe/H]} = -3.3$ in
their sample). At even lower metallicities, seven of the eight stars
known with $\mathrm{[Fe/H]} < -4.5$: SMSS~J0313$-$6708
($\mathrm{[Fe/H]} \le -7.8$, \citealt{keller2014,bessell2015});
HE~1327$-$2326 ($\mathrm{[Fe/H]} = -5.7$, \citealt{frebel2006,
aoki2006}), HE~0107$-$5240 ($\mathrm{[Fe/H]} = -5.4$,
\citealt{christlieb2004}), SDSS~J1313$-$0019 ($\mathrm{[Fe/H]} = -5.0$,
\citealt{allendeprieto2015,frebel2015}), HE~0557$-$4840
($\mathrm{[Fe/H]} = -4.8$, \citealt{norris2007}), SDSS~J1742$+$2531
($\mathrm{[Fe/H]} = -4.8$, \citealt{bonifacio2015}), SDSS~J1029$+$1729
($\mathrm{[Fe/H]} = -4.7$, \citealt{caffau2011}), and HE~0233$-$0343
($\mathrm{[Fe/H]} = -4.7$, \citealt{hansen2014}) are CEMP-no stars. The
star SDSS~J035$+$0641 may be added to this list in the near future
(\citealt{bonifacio2015} report $\mathrm{[Ca/H]} = -5.0$). The lone
exception is SDSS~J1029$+$1729, for which \citet{caffau2011} report
$\mathrm{[C/Fe]} \leq +0.9$, but higher SNR data is required in order to
be certain of its status.}\\ 

\item{{\bf The sequence of the CEMP-no and EMP $r$-II stars}: As noted above, 
the composition of the (mostly single) CEMP-no stars is dominated by
carbon, while the iron-peak elements are very weak ($\mathrm{[Fe/H]} <
-3.0$ to $< -3.5$). The strongly $r$-process-element enhanced ($r$-II) stars
appear in a relatively narrow metallicity range near $\mathrm{[Fe/H]} = -3.0$
and are mostly single, while the less-enhanced (i.e., more iron-rich) $r$-I
stars have higher metallicities. However, the extreme $r$-II stars
\object{CS~31082$-$001} and \object{HE~1523-0901}, which are not C-rich, have
absolute age determinations of $\sim$ 13-14 Gyrs from direct radioactive U/Th
chronology \citep{hill2002,frebel2007b}, so one would reasonably expect the
CEMP-no stars to be even older, at least in a chemical sense.}\\ 

\item{{\bf The bimodal distribution of $A$(C) for CEMP stars}: As discussed
above, the recent recognition that the absolute carbon abundance of CEMP
stars is apparently bimodal clearly indicates that a source of carbon
production other than that associated with AGB stars is required in the
early Universe.}\\

\item{{\bf The Li abundances of CEMP-no stars}:  As discussed by
\citet{hansen2014} and references therein, the observed abundances of
lithium for CEMP-no stars are {\it all} below the Spite Li plateau.
While many of these stars may have had their Li depleted due to internal
mixing during giant-branch evolution, this does not apply to all cases
(HE~1327$-$2326, for example, is a warm sub-giant with very low Li;
several other such stars are listed in \citealt{masseron2012}). This
provides support for the suggestion by \citet{piau2006} that Li
astration by the progenitors associated with the production of carbon in
the CEMP-no stars, followed by mixing with primordial gas, may well be
involved.}\\

\item{{\bf The observed Be and B abundance limits for BD$+$44$^\circ$493}:  
The elements Be and B are thought to form in the early Universe
exclusively by spallation reactions involving high-energy cosmic rays
\citep{prantzos2012}, which implies that the abundances of these elements in CEMP-no
stars should be uniformly low if they are indeed second-generation stars
(due to the lack of a significant background cosmic-ray flux at these
early times). \citet{placco2014b} indeed reported very low upper limits
for the abundances of Be and B in BD$+$44$^\circ$493 ($\log\epsilon$
(Be) $< -2.3$ and $\log\epsilon$ (B) $ < -0.7$). A low upper limit for
Be ($\log\epsilon$ (Be) $ < -1.8$) was also previously found by
\citet{ito2013} for this star. Although future such observations
(from the ground for Be, from space for B) are required for additional
CEMP-no stars, the results for BD$+$44$^\circ$493 are already
compelling.}\\

\item{{\bf The association of CEMP-no stars with the outer-halo population of the
Galaxy}: \citet{carollo2012} confirmed the early suggestion by Frebel et
al. (2006; see also Beers et al., in prep.) that the fraction of CEMP
stars increases with distance from the Galactic plane. The Carollo et
al. study also showed a significant contrast in the frequency of CEMP
stars between the inner- and outer-halo components of the Milky Way,
with the outer halo having roughly twice the fraction of CEMP stars as
the inner halo. They interpreted this as an indication that the
progenitor population(s) of the outer halo likely had additional
astrophysical sources of carbon production, beyond the AGB sources that
may dominate for inner-halo stars. 

Subsequently, \citet{carollo2014}
offered evidence that the CEMP-$s$ stars are preferentially associated
with the inner-halo population, while the CEMP-no stars appear more
strongly associated with the outer-halo population.  
As pointed out by \citet{carollo2014}, recent hierarchical galaxy-formation
simulations suggest that the inner-halo population of the Milky Way
arose from the assembly of relatively more-massive sub-galactic
fragments (capable of supporting extended star formation, hence reaching
higher metallicities), while the outer-halo population comprises stars
formed in relatively less-massive fragments (which experienced short or
truncated star-formation histories, and hence produced stars of lower
metallicity). This, and other lines of evidence (e.g., the likely
flatter IMF associated with star formation in the early Universe;
\citealt{tumlinson2007}) suggests that the dominant progenitors of CEMP 
stars in the two halo components were different; massive stars for the outer halo, 
and intermediate-mass stars in the case of the inner halo.}\\

\item{{\bf The discovery of damped Ly-alpha systems with enhanced carbon}:
\citet{cooke2011,cooke2012} have reported on recently discovered
high-redshift carbon-enhanced damped Ly-alpha systems that exhibit 
elemental-abundance patterns which resemble those from that expected
from massive, carbon-producing first stars, and speculated that these
progenitors are the same as those responsible for the abundance patterns
associated with CEMP-no stars in the Galaxy. To our knowledge, this is
one of the first cases, if not {\it the} first, of evidence for a direct linkage
between the observed abundances in cosmologically distant objects with
local extremely metal-poor stars. It also underscores our main conclusion that
the excess carbon was not provided by a binary companion, but was produced
elsewhere and transported across interstellar distances through the early ISM.} 

\end{itemize}

\section{Conclusions and outlook}

We have systematically monitored the radial velocities for a sample of
24 CEMP-no stars for time spans of up to eight years, and confidently
identify a total of four binary systems. We have identified a number of
other CEMP-no stars in the literature, not included in our own
programme, which similarly exhibit no signs of binarity. These include
the hyper metal-poor star HE~0107$-$5240, with RV$_{\rm mean}$ = 44.5
km~s$^{-1}$ and $\sigma = 0.3$ km~s$^{-1}$ (31 measurements from 2001 to 
2006; N. Christlieb priv. comm), the ultra metal-poor star
HE~0557$-$4840, with RV$_{\rm mean}$ = 211.7 km~s$^{-1}$ and $\sigma =
0.2$ km~s$^{-1}$ over 2193 days \citep{norris2007,
norris2012}, and the EMP star CD $-$24$^\circ$17504, with RV$_{\rm mean}$
= 136.6 km~s$^{-1}$ and $\sigma = 0.7$ km~s$^{-1}$ over 4739 days 
\citep{norris2001,aoki2009,jacobson2015b}. All of these stars have carbon 
abundances on the low-C band, whereas the majority of our detected
binary stars have carbon abundances either in the transition area between the
two bands, an area populated by both CEMP-no and CEMP-$s$ stars, or on
the high-C band. 

There clearly is a need for additional RV-monitoring observations of
CEMP-no (and CEMP-$s$) stars (see Paper~III of this series), and
detailed abundance analyses of these stars, in order to identify more
examples of high/low carbon-band associations, and to better understand
the astrophysical implications. The ongoing survey by Placco et al. to
detect bright CEMP targets from among RAVE stars with $\mathrm{[Fe/H]} <
-2$ \citep{kordopatis2013} will be an ideal source of candidates for
radial-velocity monitoring, once they can be confirmed as CEMP-no or
CEMP-$s$ stars. In addition, stars emerging from surveys such as
SkyMapper \citep{keller2007,keller2014,jacobson2015a}, the TOPoS survey
of SDSS/SEGUE turnoff stars \citep{caffau2013,bonifacio2015}, and LAMOST
\citep{deng2012,li2015} will help to further populate and confirm the
existence of the two carbon bands. 

We conclude that there now exists compelling evidence that the CEMP-no stars
are among the first low-mass stars to form in the early Universe, and as
such they contain the chemical imprints of the very first stars in their
natal clouds in the early ISM. The prime candidates for production of
large amounts of carbon in the early Universe are faint supernovae with
mixing and fallback \citep{umeda2003,nomoto2013} and massive, rapidly
rotating metal-free stars, the so-called ``spinstars'' \citep{meynet2006,
hirschi2007, maeder2015}. 

Thus far, no clear distinction between the abundance patterns of these
two scenarios has been found, either in the models or in the derived
abundances of the CEMP-no stars. The observed large over-abundances
of carbon and nitrogen are well-reproduced by both models. The spinstar
model can also explain other signatures found in CEMP-no stars, such as
low $^{12}$C/$^{13}$C isotopic ratios, and high Na, Mg, Al and Sr
abundances \citep{maeder2015}, while \citet{tominaga2014} was able to
fit the abundance patterns of 12 CEMP-no stars, from C to to Zn, with
yields from mixing and fallback SNe models. 

However, at present none of the suggested progenitors can explain the
full range of elements for which abundances are derived for CEMP-no
stars -- the light-elements, the $\alpha$-elements, the iron-peak
elements, and the neutron-capture elements. Hence, as suggested by
\citet{takahashi2014}, the observed abundance patterns of CEMP-no stars
(perhaps in particular those in the transition area between the low-C
and high-C bands) could arise from a combination of the two suggested
progenitors, or another primordial source that has yet to be
identified. Detailed abundance analyses of larger samples of CEMP-no stars
will help to settle this question.

\begin{acknowledgements}

We thank several NOT staff members and students for obtaining most of the FIES
observations for us in service mode during this large project. 
We thank Piercarlo Bonifacio and Monique Spite for sharing their
previously obtained spectrum of CS~29527-015, and obtaining a new upper
limit on the $\mathrm{[Ba/Fe]}$ ratio for this star, George 
Preston for very kindly sharing his own observations of CS~22957$-$027, and
Norbert Christlieb for sharing his RV data for HE~0107$-$5240 with us.We also
thank Heather Jacobsen for advice on the best available historical RV data for
CD-24$^\circ$17504, and David Yong for helping to clarify some of the mystery
regarding HE~1506$-$0113. Furthermore, we express our cordial thanks to the
referee for an incisive and helpful report, which led to substantial
improvements  in the paper.

T.T.H. was supported during this work by
Sonderforschungsbereich SFB 881 ``The Milky Way System'' (subproject A4)
of the German Research Foundation (DFG). J.A. and B.N. gratefully
acknowledge financial support from the Danish Natural Science Research
Council and the Carlsberg Foundation, and T.C.B., V.M.P., and J.Y. acknowledge partial
support for this work from grants PHY 08-22648; Physics Frontier
Center/{}Joint Institute or Nuclear Astrophysics (JINA), and PHY
14-30152; Physics Frontier Center/{}JINA Center for the Evolution of the
Elements (JINA-CEE), awarded by the US National Science Foundation. 

This paper is based primarily on observations made with the Nordic
OpticalTelescope, operated by the Nordic Optical Telescope Scientific 
Association at the Observatorio del Roque de los Muchachos, La Palma,
Spain, of the Instituto de Astrofisica de Canarias. 
It is also based on observations obtained at the Southern Astrophysical
Research (SOAR) telescope, which is a joint project of the Minist\'erio da
Ci\^encia, Tecnologia, e Inova\c{c}\~ao (MCTI) da Rep\'ublica Federativa do Brasil, the
U.S. National Optical Astronomy Observatory (NOAO), the University of North
Carolina at Chapel Hill (UNC), and Michigan State University (MSU).

\end{acknowledgements}


\clearpage

\appendix

\section{Heliocentric radial velocities measured for the programme stars}

\begin{table}[ht]                   
\caption{HE~0020$-$1741}         
\centering  
\begin{tabular}{lcc}
\hline\hline
HJD & RV & RV$_{err}$  \\
    & km~s$^{-1}$ & km~s$^{-1}$\\
\hline
2456191.553850 &  92.761 & 0.061\\
2456277.325704 &  92.678 & 0.085\\
2456530.727938 &  93.249 & 0.068\\
2456545.641765 &  93.144 & 0.053\\
2456603.431706 &  92.993 & 0.073\\
2456893.629396 &  93.214 & 0.062\\
2456986.461205 &  93.201 & 0.076\\
2457225.690882 &  92.920 & 0.114\\
2457257.705129 &  93.192 & 0.093\\
\hline
\end{tabular}           
\end{table}            

\begin{table}[ht]                   
\caption{CS~29527$-$015}         
\centering                       
\begin{tabular}{lcc}
\hline\hline
HJD & RV & RV$_{err}$  \\
    & km~s$^{-1}$ & km~s$^{-1}$\\
\hline
2456193.657074  & 47.032 & 1.730\\
2456214.489204  & 46.880 & 0.500\\
2456531.626981  & 47.136 & 0.083\\
2456603.479383  & 46.536 & 1.830\\
2456895.609724  & 47.777 & 1.405\\
2457257.602490  & 47.371 & 1.314\\
\hline                  
\end{tabular}           
\end{table}            

\begin{table}[ht]
\caption{CS~22166$-$016}         
\centering                       
\begin{tabular}{lcc}
\hline\hline
HJD & RV & RV$_{err}$  \\
    & km~s$^{-1}$ & km~s$^{-1}$\\
\hline
2456191.570891 & $-$210.289 & 0.220\\
2456300.388000 & $-$210.344 & 0.242\\
2456531.654302 & $-$210.477 & 0.161\\
2456579.603014 & $-$210.404 & 0.183\\
2456652.456871 & $-$209.670 & 0.903\\
2456895.638248 & $-$210.397 & 0.036\\
2456956.601783 & $-$212.384 & 1.018\\
2457225.706367 & $-$210.071 & 0.269\\
\hline
\end{tabular}           
\end{table}            

\begin{table}                   
\caption{HE~0219$-$1739}         
\centering                       
\begin{tabular}{lcc}
\hline\hline
HJD & RV & RV$_{err}$  \\
    & km~s$^{-1}$ & km~s$^{-1}$\\
\hline
2454396.611647 & 102.993 & 0.064\\
2454705.669847 & 101.817 & 0.056\\
2455126.606616 & 106.873 & 0.057\\
2455171.488102 & 107.500 & 0.056\\
2455232.347099 & 107.317 & 0.051\\
2455439.645440 & 112.468 & 0.041\\
2455485.610217 & 114.309 & 0.046\\
2455531.495382 & 113.868 & 0.063\\
2455796.705307 & 112.387 & 0.037\\
2455882.508124 & 109.832 & 0.065\\
2455945.426690 & 107.516 & 0.151\\
2456193.705723 & 102.161 & 0.052\\
2456308.451562 & 100.074 & 0.087\\
2456578.607672 & 100.814 & 0.052\\
2456603.608463 & 100.413 & 0.082\\
\hline
\end{tabular}           
\end{table}            

\begin{table}                   
\caption{BD$+$44$^\circ$493}         
\centering                       
\begin{tabular}{lcc}
\hline\hline
HJD & RV & RV$_{err}$  \\
    & km~s$^{-1}$ & km~s$^{-1}$\\
\hline
2455796.731150 & $-$150.191 & 0.019\\
2455821.602481 & $-$150.061 & 0.016\\
2455859.544117 & $-$150.050 & 0.017\\
2455882.390816 & $-$150.150 & 0.113\\
2455892.414452 & $-$150.114 & 0.125\\
2455903.488584 & $-$150.080 & 0.032\\
2455915.503192 & $-$149.985 & 0.141\\
2455971.385310 & $-$150.001 & 0.041\\
2456163.698695 & $-$150.101 & 0.164\\
2456191.605915 & $-$150.126 & 0.018\\
2456307.434407 & $-$150.041 & 0.028\\
2456518.689701 & $-$150.104 & 0.031\\
2456528.699713 & $-$150.081 & 0.028\\
2456603.638647 & $-$150.066 & 0.025\\
2456685.418052 & $-$150.040 & 0.060\\
2456893.598206 & $-$150.127 & 0.032\\
2456987.491755 & $-$150.101 & 0.086\\
2457094.349390 & $-$150.100 & 0.029\\
\hline                    
\end{tabular}             
\end{table}               

\clearpage

\begin{table}                   
\caption{HE~0405$-$0526}         
\centering                       
\begin{tabular}{lcc}
\hline\hline
HJD & RV & RV$_{err}$  \\
    & km~s$^{-1}$ & km~s$^{-1}$\\
\hline
2456190.761992 & 165.656 & 0.014\\
2456209.746333 & 165.593 & 0.057\\
2456213.772353 & 165.684 & 0.032\\
2456241.709356 & 165.674 & 0.057\\
2456340.434828 & 165.568 & 0.110\\
2456528.683688 & 165.668 & 0.029\\
2456603.556421 & 165.639 & 0.017\\
2456685.438636 & 165.676 & 0.019\\
2456686.360900 & 165.708 & 0.021\\
2456726.353768 & 165.667 & 0.071\\
2456986.480120 & 165.662 & 0.024\\
2457018.579283 & 165.700 & 0.034\\
2457094.366572 & 165.641 & 0.023\\
\hline                  
\end{tabular}           
\end{table}            


\begin{table}                   
\caption{HE~1012$-$1540}         
\centering                       
\begin{tabular}{lcc}
\hline\hline
HJD & RV & RV$_{err}$  \\
    & km~s$^{-1}$ & km~s$^{-1}$\\
\hline
2456340.537113 & 225.915 & 1.827\\
2456426.386869 & 225.897 & 0.373\\
2456458.405646 & 225.707 & 1.198\\
2456752.448580 & 226.178 & 0.180\\
2456796.400824 & 226.154 & 0.296\\
2456987.772426 & 226.366 & 0.536\\
2457076.586146 & 226.040 & 2.439\\
2457142.415352 & 226.161 & 0.349\\
\hline
\end{tabular}           
\end{table}            

\begin{table}                   
\caption{HE~1133$-$0555}         
\centering                       
\begin{tabular}{lcc}
\hline\hline
HJD & RV & RV$_{err}$  \\
    & km~s$^{-1}$ & km~s$^{-1}$\\
\hline
2454951.480644 &  270.638 & 0.226\\
2455207.661330 &  270.604 & 0.173\\
2455738.452327 &  271.020 & 0.561\\
2456005.578838 &  270.387 & 0.450\\
2456033.538287 &  269.999 & 0.599\\
2456307.743507 &  270.534 & 0.393\\
2456426.426846 &  270.804 & 0.159\\
2456796.484997 &  271.130 & 0.328\\
2457168.502711 &  270.575 & 0.537\\
\hline                   
\end{tabular}            
\end{table}

\begin{table}                   
\caption{HE~1150$-$0428}         
\centering                       
\begin{tabular}{lcc}
\hline\hline
HJD & RV & RV$_{err}$  \\
    & km~s$^{-1}$ & km~s$^{-1}$\\
\hline
2454254.469526 &  59.198 & 0.631\\
2454464.774604 &  50.506 & 0.305\\
2454481.794575 &  53.500 & 0.493\\
2454909.540207 &  43.574 & 0.403\\
2454930.649908 &  39.265 & 0.104\\
2455207.697467 &  41.831 & 0.220\\
2455704.477220 &  58.205 & 0.053\\
2456005.672746 &  56.305 & 0.094\\
2456064.445022 &  44.939 & 0.506\\
2456278.762138 &  58.039 & 0.192\\
2456399.466063 &  36.305 & 1.247\\
2456426.466103 &  37.280 & 0.053\\
2456474.416329 &  45.598 & 0.078\\
\hline                  
\end{tabular}           
\end{table}            

\begin{table}                   
\caption{HE~1201$-$1512}         
\centering                       
\begin{tabular}{lcc}
\hline\hline
HJD & RV & RV$_{err}$  \\
    & km~s$^{-1}$ & km~s$^{-1}$\\
\hline
2456722.550471 &  242.676 & 2.665\\
2456756.503043 &  238.280 & 2.525\\
2456814.436159 &  238.593 & 1.086\\
2457110.570965 &  238.337 & \dots\\
2457142.522877 &  239.365 & 0.705\\
\hline                   
\end{tabular}            
\end{table}


\begin{table}                   
\caption{HE~1300$+$0157}         
\centering                       
\begin{tabular}{lcc}
\hline\hline
HJD & RV & RV$_{err}$  \\
    & km~s$^{-1}$ & km~s$^{-1}$\\
\hline
2456756.542595 &  75.397 & 0.531\\
2456796.569354 &  74.182 & 0.418\\
2456840.398240 &  75.141 & 0.321\\
2457142.604191 &  73.800 & 0.284\\
2457168.543909 &  74.159 & 0.146\\
\hline                  
\end{tabular}           
\end{table}            


\begin{table}                   
\caption{BS~16929$-$005}         
\centering                       
\begin{tabular}{lcc}
\hline\hline
HJD & RV & RV$_{err}$  \\
    & km~s$^{-1}$ & km~s$^{-1}$\\
\hline
2456340.583325 & $-$50.188 & 2.672\\
2456426.503527 & $-$51.478 & 0.973\\
2456652.666926 & $-$50.625 & 1.020\\
2456722.615544 & $-$50.970 & 1.810\\
2456756.432118 & $-$49.880 & 1.415\\
2457096.737115 & $-$50.575 & 0.166\\
2457225.402088 & $-$51.430 & 0.814\\
\hline                  
\end{tabular}           
\end{table}            

\begin{table}                   
\caption{HE~1300$-$0641}         
\centering                       
\begin{tabular}{lcc}
\hline\hline
HJD & RV & RV$_{err}$  \\
    & km~s$^{-1}$ & km~s$^{-1}$\\
\hline
2456756.628997 &  68.926 & 0.893\\
2457142.577242 &  68.765 & 0.405\\
\hline
\end{tabular}           
\end{table}            

\begin{table}                   
\caption{HE~1302$-$0954}         
\centering                       
\begin{tabular}{lcc}
\hline\hline
HJD & RV & RV$_{err}$  \\
    & km~s$^{-1}$ & km~s$^{-1}$\\
\hline
2456756.590880 &  32.494 & 0.268\\
2457111.063422 &  32.549 & 0.229\\
2457142.552710 &  32.570 & 0.124\\
\hline
\end{tabular}           
\end{table}            

\clearpage

\begin{table}                   
\caption{CS~22877$-$001}         
\centering                       
\begin{tabular}{lcc}
\hline\hline
HJD & RV & RV$_{err}$  \\
    & km~s$^{-1}$ & km~s$^{-1}$\\
\hline
2454219.526768 & 166.288 & 0.071\\
2454254.500103 & 166.360 & 0.075\\
2454909.559431 & 166.422 & 0.087\\
2454951.514264 & 166.194 & 0.070\\
2454964.525457 & 166.332 & 0.082\\
2455207.715279 & 166.330 & 0.074\\
2455344.525470 & 166.207 & 0.064\\
2455554.795730 & 166.266 & 0.044\\
2455620.725908 & 166.220 & 0.058\\
2455704.535965 & 166.112 & 0.045\\
2456005.692575 & 166.283 & 0.087\\
2456033.654814 & 166.519 & 0.091\\
2456033.711811 & 166.402 & 0.247\\
2456722.657147 & 166.376 & 0.152\\
2457142.537184 & 166.147 & 0.082\\
\hline
\end{tabular}           
\end{table}            


\begin{table}                   
\caption{HE~1327$-$2326}         
\centering                       
\begin{tabular}{lcc}
\hline\hline
HJD & RV & RV$_{err}$  \\
    & km~s$^{-1}$ & km~s$^{-1}$\\
\hline
2454219.542926 &   65.578 & \dots\\
2454930.612715 &   64.974 & \dots\\
2455232.754648 &   62.764 & \dots\\
2455344.464871 &   63.630 & \dots\\
2455620.712669 &   65.347 & \dots\\
2455704.502164 &   64.463 & 0.935\\
2455945.751608 &   62.329 & \dots\\
2456006.750000 &   65.188 & \dots\\
2456796.598655 &   64.818 & 0.159\\
\hline
\end{tabular}   
\tablefoot{Only the spectral order containing the Mg triplet could be used
  for the correlation for the majority of the spectra of this
  star, thus an internal error could not be calculated for these spectra.}        
\end{table}            


\begin{table}                   
\caption{HE~1410$+$0213}         
\centering                       
\begin{tabular}{lcc}
\hline\hline
HJD & RV & RV$_{err}$  \\
    & km~s$^{-1}$ & km~s$^{-1}$\\
\hline
2454219.643865 &  80.983 & 0.024\\
2454314.427327 &  81.213 & 0.028\\
2454516.758372 &  81.111 & 0.159\\
2454909.570456 &  81.069 & 0.046\\
2454930.734724 &  81.026 & 0.036\\
2454951.686449 &  80.874 & 0.028\\
2455232.768813 &  81.025 & 0.020\\
2455262.603235 &  80.918 & 0.056\\
2455344.565415 &  81.068 & 0.028\\
2455555.796783 &  81.291 & 0.178\\
2455620.765457 &  80.837 & 0.030\\
2455662.717240 &  80.960 & 0.021\\
2455704.599755 &  81.183 & 0.021\\
2455776.420181 &  81.414 & 0.016\\
2456005.703435 &  81.143 & 0.042\\
2456033.595529 &  81.330 & 0.043\\
2456078.490141 &  81.348 & 0.032\\
2456722.586745 &  81.417 & 0.173\\
2457079.771538 &  81.389 & 0.055\\
2457096.777330 &  81.357 & 0.050\\
2457110.655228 &  81.202 & 0.103\\
2457142.633878 &  81.069 & 0.031\\
2457225.423528 &  80.984 & 0.037\\
\hline
\end{tabular}           
\end{table}            

\begin{table}                   
\caption{HE~1506$-$0113}         
\centering                       
\begin{tabular}{lcc}
\hline\hline
HJD & RV & RV$_{err}$  \\
    & km~s$^{-1}$ & km~s$^{-1}$\\
\hline
2456756.666987 & $-$85.865 & 1.202\\
2456796.621783 & $-$85.830 & 0.153\\
2456813.637005 & $-$83.626 & 0.952\\
2456893.378926 & $-$83.223 & 0.594\\
2457142.677901 & $-$81.052 & 0.524\\
2457190.528825 & $-$79.152 & 0.449\\
2457225.469049 & $-$77.545 & 0.943\\
2457239.418486 & $-$80.405 & 1.765\\
2457241.405630 & $-$80.661 & 0.736\\
2457249.435833 & $-$79.985 & 0.798\\
\hline                  
\end{tabular}           
\end{table}            


\begin{table}                   
\caption{CS~22878$-$027}         
\centering                       
\begin{tabular}{lcc}
\hline\hline
HJD & RV & RV$_{err}$  \\
    & km~s$^{-1}$ & km~s$^{-1}$\\
\hline
2456191.343802 &  $-$91.097 & 0.810\\
2456756.724060 &  $-$89.769 & 0.331\\
2456796.698016 &  $-$90.696 & 1.839\\
2456887.392049 &  $-$90.533 & 1.495\\
2457142.721805 &  $-$90.845 & 1.446\\
2457168.690235 &  $-$92.330 & 1.021\\
2457225.500502 &  $-$90.822 & 2.506\\
\hline                   
\end{tabular}           
\end{table}

\clearpage

\begin{table}                   
\caption{CS~29498$-$043}         
\centering                       
\begin{tabular}{lcc}
\hline\hline
HJD & RV & RV$_{err}$  \\
    & km~s$^{-1}$ & km~s$^{-1}$\\
\hline
2454314.535743 & $-$32.557 & 1.583\\
2454338.491418 & $-$32.080 & 0.964\\
2454373.380034 & $-$32.911 & 1.247\\
2454665.594229 & $-$32.885 & 1.988\\
2454705.541235 & $-$33.060 & 0.421\\
2455059.460229 & $-$32.215 & 1.468\\
2455070.509288 & $-$33.641 & 2.553\\
2455149.322697 & $-$31.798 & 1.008\\
2455415.540797 & $-$32.527 & 1.318\\
2455439.469033 & $-$32.242 & 1.385\\
2455738.656324 & $-$31.819 & 1.760\\
2455776.589351 & $-$32.193 & 0.682\\
2456033.722001 & $-$31.172 & 0.865\\
2456458.700787 & $-$32.383 & 1.420\\
2456917.501102 & $-$33.836 & 1.839\\
\hline
\end{tabular}           
\end{table}            


\begin{table}                   
\caption{CS~29502$-$092}         
\centering                       
\begin{tabular}{lcc}
\hline\hline
HJD & RV & RV$_{err}$  \\
    & km~s$^{-1}$ & km~s$^{-1}$\\
\hline
2454314.615516 & $-$67.045 & 0.061\\
2454373.458720 & $-$67.215 & 0.051\\
2454625.669705 & $-$67.200 & 0.061\\
2454665.647386 & $-$67.284 & 0.060\\
2454705.583752 & $-$67.143 & 0.054\\
2454964.724327 & $-$67.152 & 0.071\\
2455059.495880 & $-$67.266 & 0.065\\
2455126.513232 & $-$67.213 & 0.088\\
2455174.330130 & $-$67.225 & 0.055\\
2455344.673757 & $-$67.116 & 0.068\\
2455415.485690 & $-$67.193 & 0.066\\
2455439.439882 & $-$67.216 & 0.051\\
2455503.319269 & $-$67.452 & 0.088\\
2455531.397775 & $-$67.351 & 0.105\\
2455704.690065 & $-$67.173 & 0.082\\
2455776.526490 & $-$67.174 & 0.055\\
2456139.661574 & $-$67.368 & 0.066\\
2456191.367411 & $-$67.157 & 0.073\\
2456458.712569 & $-$67.180 & 0.068\\
2456917.518394 & $-$67.175 & 0.100\\
\hline                  
\end{tabular}           
\end{table}            

\begin{table}                   
\caption{HE~2318$-$1621}         
\centering                       
\begin{tabular}{lcc}
\hline\hline
HJD & RV & RV$_{err}$  \\
    & km~s$^{-1}$ & km~s$^{-1}$\\
\hline
2456191.508735 & $-$41.724 & 0.136\\
2456458.678220 & $-$41.780 & 0.085\\
2456512.729591 & $-$41.299 & 0.228\\
2456574.601390 & $-$42.136 & 0.210\\
2456888.556884 & $-$41.612 & 0.299\\
2456956.516288 & $-$41.446 & 0.270\\
2457225.639412 & $-$41.891 & 0.112\\
\hline                  
\end{tabular}           
\end{table}            

\begin{table}                   
\caption{CS~22949$-$037}         
\centering                       
\begin{tabular}{lcc}
\hline\hline
HJD & RV & RV$_{err}$  \\
    & km~s$^{-1}$ & km~s$^{-1}$\\
\hline
2456191.418846 & $-$125.488 & 0.490\\
2456213.439688 & $-$125.619 & 0.423\\
2456518.588656 & $-$125.438 & 1.300\\
2456574.639061 & $-$126.021 & 1.465\\
2456881.545231 & $-$125.238 & 1.834\\
2456886.533402 & $-$125.775 & 1.247\\
2456956.499023 & $-$125.341 & 1.153\\
\hline
\end{tabular}           
\end{table}


\begin{table}                   
\caption{CS~22957$-$027}         
\centering                       
\begin{tabular}{lcc}
\hline\hline
HJD & RV & RV$_{err}$  \\
    & km~s$^{-1}$ & km~s$^{-1}$\\
\hline
2454314.650763 & $-$61.056 & 0.053\\
2454373.602779 & $-$61.598 & 0.071\\
2454396.522335 & $-$62.847 & 0.209\\
2454480.353018 & $-$63.679 & 0.119\\
2454665.720088 & $-$72.551 & 0.112\\
2454780.480070 & $-$76.573 & 0.109\\
2454819.304024 & $-$76.207 & 0.170\\
2455059.670725 & $-$67.280 & 0.056\\
2455071.664734 & $-$67.767 & 0.388\\
2455126.567510 & $-$65.441 & 0.085\\
2455149.455854 & $-$64.285 & 0.130\\
2455207.320651 & $-$63.146 & 0.131\\
2455439.578111 & $-$61.418 & 0.084\\
2455503.374442 & $-$62.705 & 0.525\\
2455531.350698 & $-$62.359 & 0.139\\
2455738.721957 & $-$72.151 & 0.063\\
2455776.669030 & $-$73.957 & 0.105\\
2455882.450486 & $-$76.464 & 0.111\\
\hline                  
\end{tabular}           
\end{table}

\clearpage

\section{Literature data for the single programme stars}

\begin{table}[ht]
\caption{Mean heliocentric radial velocities from the literature and total
  time-span covered for the single stars }
\label{tbl-B1}
\centering
\begin{tabular}{lrrrrl}
\hline\hline
Stellar ID & $\Delta$T Total &  mean RV (this work)  & mean RV & N & Ref \\
           & (days)          &(km~s$^{-1}$)          & (km~s$^{-1}$)&   &  \\
\hline
\object{HE~0020$-$1741}& 3333 &$+$93.018 &$+$93.0 &1 &\citet{kordopatis2013} \\
\hline
\object{CS~29527$-$015}& 7745 &$+$47.077 &$+$45.7  &1 &\citet{norris1996}\\ 
                       &      &          &$+$48.0  &1 &\citet{aoki2013} \\
\hline
\object{CS~22166$-$016}& 3727 &$-$209.769&$-$210.0 &1 & \citet{roederer2014}\\
\hline
\object{BD$+$44$^\circ$493}&11368&$-$150.084&$-$151.3&6& \citet{carney1986}   \\
                       &      &          &$-$150.6 &28& \citet{carney2003} \\
                       &      &          &$-$150.3 &4 & \citet{ito2013}\\
                       &      &          &$-$150.0 &2 & \citet{roederer2014}\\
                       &      &          &$-$150.1 &4 & \citet{starkenburg2014}\\
\hline
\object{HE~1012$-$1540}& 4646 &$+$226.362&$+$225.6 &2 &\citet{cohen2008}\\
                       &      &          &$+$226.3 &2 & \citet{cohen2013}\\
                       &      &          &$+$225.6 &2 & \citet{roederer2014} \\
\hline
\object{HE~1201$-$1512}& 2524 &$+$239.450&$+$238.0 &4 & \citet{norris2013a}\\
\hline
\object{HE~1300$+$0157}& 4338 &$+$74.536 &$+$74.3  &3 & \citet{barklem2005}\\
                       &      &          &$+$73.4  &1 & \citet{cohen2008}\\
                       &      &          &$+$74.5  &6 & \citet{starkenburg2014}\\  
\hline
\object{BS~16929$-$005}& 5157 &$-$50.619 &$-$51.2  &1& \citet{honda2004}\\
                       &      &          &$-$54.0  &1& \citet{lai2004}\\
                       &      &          &$-$50.4  &1& \citet{aoki2007}\\
                       &      &          &$-$51.7  &1& \citet{lai2008}\\
                       &      &          &$-$50.5  &5&\citet{starkenburg2014}\\
\hline
\object{HE~1300$-$0641}& 4329 &$+$68.822 &$+$67.9  &2 & \citet{barklem2005}\\
\hline
\object{CS~22877$-$001}&6228 &$+$166.297&$+$166.1 &1 &\citet{aoki2002a}\\
\hline
\object{HE~1327$-$2326}& 3660 &$+$64.344 &$+$63.9  &1 & \citet{aoki2006}\\
\hline
\object{HE~1410$+$0213}& 4929 &$+$81.140 &$+$80.7  &1 & \citet{cohen2013}\\
\hline
\object{CS~22878$-$027}& 4355 &$-$91.016 &$-$91.3  &2 & \citet{lai2008}\\
                       &      &          &$-$91.2  &2 & \citet{roederer2014}\\
                       &      &          &$-$91.5  &6 & \citet{starkenburg2014}\\
\hline
\object{CS~29498$-$043}& 4800 &$-$32.488 &$-$32.5  &1 & \citet{aoki2002b}\\
                       &      &          &$-$32.6  &1 & \citet{aoki2002c}\\
                       &      &          &$-$32.9  &3 & \citet{aoki2004}\\
                       &      &          &$-$32.6  &2 & \citet{roederer2014}\\

\hline
\object{CS~29502$-$092}& 4782 &$-$67.215 &$-$67.7  &1 &\citet{tsangarides2003}\\
                       &      &          &$-$67.0  &1 &\citet{lai2004}\\
                       &      &          &$-$65.2  & 1&\citet{ruchti2011}\\
                       &      &          &$-$67.0  &1 &\citet{sakari2013}\\
                       &      &          &$-$66.6  &1 &\citet{roederer2014}\\
                       &      &          &$-$66.8  &3 & \citet{starkenburg2014}\\
\hline
\object{CS~22949$-$037}& 5129 &$-$125.560&$-$126.4 &1 &\citet{mcwilliam1995}\\
                       &      &          &$-$125.7 &1 &\citet{norris2001}\\
                       &      &          &$-$125.6 &4 &\citet{depagne2002}\\ 
                       &      &          &$-$125.6 &1 &\citet{bonifacio2009}\\
                       &      &          &$-$125.4 & 2&\citet{roederer2014}\\
                       &      &          &$-$125.9 &2 &\citet{starkenburg2014}\\
\hline
\end{tabular}
\tablebib{Note that no additional radial-velocity measurements was found for
  HE~0405$-$0526, HE~1133$-$0555, HE~1302$-$0954 and HE~2318$-$1621.}
\end{table}



\begin{thebibliography}{}

\bibitem[Allende Prieto et al.(2015)]{allendeprieto2015} Allende Prieto, C.,
  Fern{\'a}ndez-Alvar, E., Aguado, D. S., et al., 2015, \aap, 579, A98

\bibitem[Abate et al.(2013)]{abate2013} Abate, C., Pols, O. R., Izzard, R. G.,
  et al., 2013, \aap, 552, A26

\bibitem[Abate et al.(2015)]{abate2015} Abate, C., Pols, O. R., Stancliffe, et
  al., 2015, \aap, 581, A62

\bibitem[Aoki et al.(2002a)]{aoki2002a} Aoki, W., Norris, J.E., Ryan, S.G.,
  Beers, T.C., \& Ando, H., 2002a, \apj, 567, 1166

\bibitem[Aoki et al.(2002b)]{aoki2002b} Aoki, W., Ando, H., Honda, S., et al.,
  2002b, \pasj, 54, 427

\bibitem[Aoki et al.(2002c)]{aoki2002c} Aoki, W., Norris, J.E., Ryan, S.G.,
  Beers, T.C., \& Ando, H., 2002c, \apjl, 576, L141

\bibitem[Aoki et al.(2004)]{aoki2004} Aoki, W., Norris, J. E., Ryan, S. G., et
  al., 2004, \apjl, 608, 971

\bibitem[Aoki et al.(2006)]{aoki2006} Aoki, W., Frebel, A., Christlieb, N., et
  al., 2006, \apj, 639, 897

\bibitem[Aoki et al.(2007)]{aoki2007} Aoki, W., Beers, T. C., Christlieb, N.,
  Norris, J. E., Ryan, S. G., \& Tsangarides, S., 2007, \apj, 655, 492

\bibitem[Aoki et al.(2009)]{aoki2009} Aoki, W., Barklem, P. S., Beers, T. C.,
  et al., 2009, \apj, 698, 1803

\bibitem[Aoki et al.(2013)]{aoki2013} Aoki, W., Beers, T. C., Lee, Y. S., et
  al., 2013, \aj, 145, 13

\bibitem[Asplund et al.(2009)]{asplund2009} Asplund, M., Grevesse, N.,
   Sauval, A. J., \& Scott, P. 2009, \araa, 47, 481

\bibitem[Barklem et al.(2005)]{barklem2005} Barklem, P.S., Christlieb, N.,
  Beers, T.C., et al., 2005, \aap, 439, 129

\bibitem[Beers et al.(1985)]{beers1985} Beers, T.C., Preston, G.W.,\&
  Shectman, S.A., 1985, \aj, 90, 2089

\bibitem[Beers et al.(1992)]{beers1992} Beers, T.C., Preston, G.W.,\&
  Shectman, S.A., 1992, \aj, 103, 1987

\bibitem[Beers et al.(2007)]{beers2007} Beers, T.C., Flynn, C., Rossi, S., et
  al., 2007, \apjs, 168, 128

\bibitem[Beers et al.(2014)]{beers2014} Beers, T. C., Norris, J. E., Placco,
  V. M., et al., 2014, \apj, 794, 58

\bibitem[Beers \& Christlieb(2005)]{beerschristlieb2005} Beers, T.C., \&
  Christlieb N., 2005, \araa, 43, 531

\bibitem[Bessell et al.(2004)]{bessell2004} Bessell, M. S., Christlieb, N., \&
  Gustafsson, B., 2004, \apjl, 612, L61

\bibitem[Bessell et al.(2015)]{bessell2015} Bessell, M. S., Collet, R.,
  Keller, S. C., et al., 2015, \apjl, 806, L16

\bibitem[Bonifacio et al.(2009)]{bonifacio2009} Bonifacio, P., Spite, M.,
  Cayrel, R., et al., 2009, \aap, 501, 519

\bibitem[Bonifacio et al.(2015)]{bonifacio2015} Bonifacio, P., Caffau, E.,
  Spite, M., et al., 2015, \aap, 579, 28

\bibitem[Bromm \& Loeb(2003)]{bromm2003} Bromm, V., \& Loeb, A., 2003, \nat,
  425, 812

\bibitem[Caffau et al.(2011)]{caffau2011} Caffau, E., Bonifacio, P., Francois,
  P., et al., 2011, \nat, 477, 67 

\bibitem[Caffau et al.(2013)]{caffau2013} Caffau, E., Bonifacio, P., Sbordone,
  L., et al., 2013, \aap, 560, A71

\bibitem[Carney \& Latham (1986)]{carney1986} Carney, B. W., \& Latham, D. W.,
  1986, \aj, 92, 60

\bibitem[Carney et al.(2003)]{carney2003} Carney, B.W., Latham, D.W.,
  Stefanik, R.P., Laird, J.B., \& Morse, J.A., 2003, \aj, 125, 293 

\bibitem[Carollo et al.(2012)]{carollo2012} Carollo, D., Beers, T. C., Bovy,
  J., et al., 2012, \apj, 744, 195

\bibitem[Carollo et al.(2014)]{carollo2014} Carollo, D., Freeman, K., Beers,
  T. C., et al., 2014, \apj, 788, 180 

\bibitem[Cayrel et al.(2004)]{cayrel2004} Cayrel, R., Depagne, E., Spite, M.,
  et al., 2004, \aap, 416, 1117

\bibitem[Christlieb et al.(2004)]{christlieb2004} Christlieb, N., Gustafsson,
  B., Korn, A.J., et al., 2004, \apj, 603, 708

\bibitem[Christlieb et al.(2008)]{christlieb2008} Christlieb, N., Sch{\"o}rck,
  T., Frebel, A., Beers, T.C., Wisotzki, L., \& Reimers, D., 2008, \aap, 484,
  721 

\bibitem[Cohen et al.(2008)]{cohen2008} Cohen, J. G., Christlieb, N.,
  McWilliam, A., et al., 2008, \apj, 672, 320

\bibitem[Cohen et al.(2013)]{cohen2013} Cohen, J.G., Christlieb, N., Thompson,
  I., et al., 2013, \apj, 778, 56

\bibitem[Cooke et al.(2011)]{cooke2011} Cooke, R., Pettini, M., Steidel,
  C. C., Rudie, G. C., \& Jorgenson, R. A., 2011, \mnras, 412, 1047

\bibitem[Cooke et al.(2012)]{cooke2012} Cooke, R., Pettini, M., \& Murphy,
  M. T., 2012, \mnras, 425, 347

\bibitem[Deng et al.(2012)]{deng2012} Deng, L. C., Newberg, H. J., Liu, C., et
  al., 2012, Research in Astronomy and Astrophysics, 12, 735

\bibitem[Depagne et al.(2002)]{depagne2002} Depagne, E., Hill, V., Spite, M.,
  et al., 2002, \aap, 390, 187

\bibitem[Duquennoy \& Mayor (1991)]{duquennoy1991} Duquennoy, A., \& Mayor, M., 1991,
  \aap, 248, 485

\bibitem[Frebel et al.(2006)]{frebel2006} Frebel, A., Christlieb, N., Norris,
  J.E., Aoki, W., \& Asplund, M., 2006, \apjl, 638, L17

\bibitem[Frebel et al.(2007a)]{frebel2007a} Frebel, A., Johnson, J. L., \&
  Bromm, V., 2007a, \mnras, 380, L40

\bibitem[Frebel et a.(2007b)]{frebel2007b} Frebel, A., Christlieb, N.,
Norris, J.E., et al. 2007b, \apj, 660, L117

\bibitem[Frebel et al.(2008)]{frebel2008} Frebel, A., Collet, R., Eriksson,
  K., Christlieb, N., \& Aoki, W., 2008, \apj, 684, 588

\bibitem[Frebel et al.(2015)]{frebel2015} Frebel, A., Chiti, A., Ji, A. P.,
  Jacobson, H. R., \& Placco, V. M., 2015, \apjl, submitted (ArXiv:1507.01973)

\bibitem[Frebel \& Norris (2015)]{frebelnorris2015} Frebel, A., \& Norris,
  J. E., 2015, \araa, 53, 631

\bibitem[Giridhar et al.(2001)]{giridhar2001} Giridhar, S., Lambert, D.L.,
  Gonzalez, G., \& Pandey, G., 2001, \pasp, 113, 519

\bibitem[Hansen et al.(2011)]{hansen2011} Hansen, T., Andersen, J., Nordstr{\"o}m, 
  B., Buchhave, L., \& Beers, T.C.: 2011, \apj, 743, L1 

\bibitem[Hansen et al.(2014)]{hansen2014} Hansen, T., Hansen, C.J.,
  Christlieb, N., et al., 2014, \apj, 787, 162
 
\bibitem[Hansen et al.(2015a)]{hansen2015a} Hansen, T., Hansen, C.J.,
  Christlieb, N., et al., 2015a, \apj, 807, 173

\bibitem[Hansen et al.(2015b)]{hansen2015b} Hansen, T., Andersen, J.,
  Nordstr{\"o}m, B., et al., 2015b, \aap, in press (Paper~I; 
  arXiv1509.05344)

\bibitem[Henden et al.(2015)]{henden2015} Henden, A.A., Levine, S., Terrell,
  D., \& Welch, D.L., 2015, American Astronomical Society Meeting Abstracts, 225, 336.16

\bibitem[Hill et al.(2002)]{hill2002} Hill, V., Plez, B., Cayrel, R., et al.,
  2002, \aap, 387, 560

\bibitem[Hirschi(2007)]{hirschi2007} Hirschi, R., 2007, \aap, 461, 571

\bibitem[Honda et al.(2004)]{honda2004} Honda, S., Aoki, W., Ando, H., et al.,
  2004, \apjs, 152, 113

\bibitem[Ito et al.(2013)]{ito2013} Ito, H., Aoki, W., Beers, T.C., Tominaga,
  N., Honda, S., \& Carollo, D., 2013, \apj, 773, 33

\bibitem[Ivezic et al.(2012)]{ivezic2012} Ivezic, Z., Beers, T.C., \& Juric, M.
2012, \araa, 50, 251

\bibitem[Izzard et al.(2009)]{izzard2009} Izzard, R. G., Glebbeek, Stancliffe,
  R. J., \& Pols, O. R., 2009, \aap, 508, 1359

\bibitem[Jacobson et al.(2015a)]{jacobson2015a} Jacobson, H., Keller, S., Frebel, A.,
et al., 2015a, \apj, 807, 171

\bibitem[Jacobson et al.(2015b)]{jacobson2015b} Jacobson, H.,\&  Frebel, A.,
  2015b, \apj, 808, 53

\bibitem[Jorissen et al.(2015)]{jorissen2015} Jorissen, A., Hansen, T., Van 
  Eck, S., et al., 2015, \aap, in press (arXiv1510.06045) 

\bibitem[Keller et al.(2007)]{keller2007} Keller, S. G., Schmidt, B. P.,
  Bessell, M. S., et al., 2007, \pasa, 24, 1

\bibitem[Keller et al.(2014)]{keller2014} Keller, S. C., Bessell, M. S., Frebel,
  A., et al., 2014, \nat, 506, 463

\bibitem[Kennedy et al.(2011)]{kennedy2011} Kennedy, C.R., Sivarani, T.,
  Beers, T.C., et al., 2011, \aj, 141, 102

\bibitem[Klessen et al.(2012)]{klessen2012} Klessen, R. S., Glover, S. C. O.,
  Clark, P. C., 2012, \mnras, 421, 3217

\bibitem[Kordopatis et al.(2013)]{kordopatis2013} Kordopatis, G., Gilmore, G.,
  Steinmetz, M., et al., 2013, \aj, 146, 134

\bibitem[Kupka et al.(1999)]{kupka1999} Kupka, F., Piskunov, N., 
Ryabchikova, T. A., Stempels, H. C., \& Weiss, W. W.\ 1999, \aaps, 138, 119 

\bibitem[Lai et al.(2004)]{lai2004} Lai, D. K., Bolte, M., Johnson, J. A., \&
  Lucatello, S., 2004, \aj, 128, 2402

\bibitem[Lai et al.(2008)]{lai2008} Lai, D K., Bolte, M., Johnson, J. A., et
  al., 2008, \apj, 681, 1524

\bibitem[Lee et al.(2013)]{lee2013} Lee, Y. S., Beers, T. C., Masseron, T., et
  al., 2013, \aj, 146, 132

\bibitem[Li et al.(2015)]{li2015} Li, H., Zhao, G., Christlieb, N., et al.,
  2015, \apj, 798, 110

\bibitem[Lucatello et al.(2005)]{lucatello2005} Lucatello, S., Tsangarides, 
   S., Beers, T. C., et al., 2005, \apj, 625, 825

\bibitem[Lucatello et al.(2006)]{lucatello2006} Lucatello, S., Beers, T. C.,
  Christlieb, N., 2006, \apjl, 652, L37

\bibitem[Maeder et al.(2015)]{maeder2015} Maeder, A., Meynet, G., \&
  Chiappini, C., 2015, \aap, 576, A56

\bibitem[Masseron et al.(2012)]{masseron2012} Masseron, T., Johnson, J.,
  Lucatello, S., et al., 2012, \apj, 751, 14

\bibitem[Mathieu et al.(1990)]{mathieu1990} Mathieu, R.D., Latham, D.W., \& Griffin,
   R.F., 1990, \aj, 100, 1899

\bibitem[Mattsson (2015)]{mattsson2015} Mattsson, L., \mnras, in press 
   (arXiv:1505.04758) 

\bibitem[McWilliam et al.(1995)]{mcwilliam1995} Mcwilliam, A., Preston, G. W.,
  Sneden, C., \& Shectman, S., 1995, \aj, 109, 2736

\bibitem[Mermilliod et al.(2007)]{mermio2007} Mermilliod, J.-C., Andersen, J.,
  Latham, D.W., \& Mayor, M., 2007, \aap, 473, 829 

\bibitem[Meynet et al.(2006)]{meynet2006} Meynet, G., Ekstr{\"o}m,
  S., \& Maeder, A., 2006, \aap, 447, 623

\bibitem[Nomoto et al.(2013)]{nomoto2013} Nomoto, K., Kobayashi, C., \&
  Tominaga, N., 2013, \araa, 51, 457 

\bibitem[Nordstr{\"o}m et al.(1997)]{nordstrom1997} Nordstr{\"o}m, B., 
  Andersen, J., \& Andersen, M.I., 1997, \aap, 322, 460 

\bibitem[Norris et al.(1996)]{norris1996} Norris, J. E., Ryan, S. G., \&
  Beers, T. C., 1996, \apjs, 107, 391

\bibitem[Norris et al.(1999)]{norris1999} Norris, J.E., Ryan, S.G., \& Beers,
  T.C., 1999, \apjs, 123, 639

\bibitem[Norris et al.(2001)]{norris2001} Norris, J. E., Ryan, S. G., \&
  Beers, T. C., 2001, \apj, 561, 1034

\bibitem[Norris et al.(2007)]{norris2007} Norris, J.E., Christlieb, N., Korn,
  A.J., et al., 2007, \apj, 670, 774

\bibitem[Norris et al.(2012)]{norris2012} Norris, J. E., Christlieb, N.,
  Bessell, M. S., Asplund, M., Eriksson, K., \& Korn, A. J., 2012, \apj, 753,
  150

\bibitem[Norris et al.(2013a)]{norris2013a} Norris, J. E., Bessell, M. S.,
  Yong, D., et al., 2013, \apj, 762, 25

\bibitem[Norris et al.(2013b)]{norris2013b} Norris, J. E., Yong, D., Bessell,
  M. S., et al., 2013b, \apj, 762, 28

\bibitem[Piau et al.(2006)]{piau2006} Piau, L., Beers, T. C., Balsara, D. S.,
  Sivarani, T., Truran, J. W., \& Ferguson, J. W., 2006, \apj, 653, 300

\bibitem[Placco et al.(2010)]{placco2010} Placco, V.M., Kennedy, C.R., Rossi,
  S., et al., 2010, \aj, 139, 1051

\bibitem[Placco et al.(2014a)]{placco2014a} Placco, V.M., Frebel, A., Beers,
  T.C., et al., 2014a, \apj, 781, 40

\bibitem[Placco et al.(2014b)]{placco2014b} Placco, V. M., Beers, T. C.,
  Roederer, I. U., et al., 2014b, \apj, 790, 34

\bibitem[Prantzos (2012)]{prantzos2012} Prantzos, N. 2012, \aap, 542, 67

\bibitem[Preston \& Sneden(2001)]{prestonsneden2001} Preston, G.W., \& Sneden,
  C., 2001, \aj, 122, 1545

\bibitem[Preston et al.(1991)]{preston1991} Preston, G.W., Shectman, S.A., \&
  Beers, T.C., 1991, \apjs, 76, 1001

\bibitem[Riebel et al.(2010)]{riebel2010} Riebel, D., Meixner, M., Fraser, O. 
   et al., \apj, 723, 1195

\bibitem[Roederer et al.(2014)]{roederer2014} Roederer I.U., Preston, G.W.,
  Thompson, I.B., 2014, \aj, 147, 136

\bibitem[Ruchti et al.(2011)]{ruchti2011} Ruchti, G. R., Fulbright, J. P.,
  Wyse, R. F. G., et al., 2011, \apj, 737, 9

\bibitem[Ryan et al.(2005)]{ryan2005} Ryan, S.G., Aoki, W., Norris, J.E.,
  \& Beers, T.C. 2005, \apj, 635, 349 

\bibitem[Sakari et al.(2013)]{sakari2013} Sakari, C. M., Shetrone, M., Venn,
  K., McWilliam, A., \& Dotter, A., 2013, \mnras, 343, 358

\bibitem[Spite et al.(2013)]{spite2013} Spite, M., Caffau, E., Bonifacio, P.,
  et al., 2013, \aap, 552, A107 

\bibitem[Starkenburg et al.(2014)]{starkenburg2014} Starkenburg, E., Shetrone,
  M.D., McConnachie, A.W., \& Venn, K.A., 2014, \mnras, 411, 1217

\bibitem[Takahashi et al.(2014)]{takahashi2014} Takahashi, K., Umeda, H., \&
  Yoshida, T., 2014, \apj, 794, 40

\bibitem[Tominaga et al.(2014)]{tominaga2014} Tominaga, N., Iwamoto, N., \&
  Nomoto, K., 2014, \apj, 785, 98

\bibitem[Tsangarides et al.(2003)]{tsangarides2003} Tsangarides, S. A., Ryan,
  S. G., \& Beers, T. C., 2003, Astronomical Society of the Pacific Conference
  Series, ed: Charbonnel, C., Schaerer, D., \& Meynet, G., 133

\bibitem[Tumlinson(2007)]{tumlinson2007} Tumlinson, J. 2007, \apj, 665, 1361

\bibitem[Umeda \& Nomoto(2003)]{umeda2003} Umeda, H., \& Nomoto, K., 2003,
  \nat, 422, 871

\bibitem[Yong et al.(2013)]{yong2013} Yong, D., Norris, J.E., Bessell, M.S.,
  et al., 2013, \apj, 762, 26

\bibitem[Wisotzki et al.(2000)]{wisotzki2000} Wisotzki, L., Christlieb, N., Bade, N., et al. 2000, \aap, 358, 77

\end{thebibliography}
\end{document}